\documentclass{article}

\usepackage{microtype}
\usepackage{graphicx}
\usepackage{subcaption}
\usepackage{booktabs}

\usepackage{hyperref}

\usepackage[preprint]{icml2026}

\usepackage{amsmath}
\usepackage{amssymb}
\usepackage{mathtools}
\usepackage{amsthm}

\usepackage[capitalize,noabbrev]{cleveref}

\theoremstyle{plain}

\theoremstyle{definition}

\theoremstyle{remark}

\usepackage[dvipsnames]{xcolor}

\usepackage{wrapfig}

\usepackage[textsize=tiny]{todonotes}

\icmltitlerunning{Canzona: A Unified, Asynchronous, and Load-Balanced Framework for Distributed Matrix-based Optimizers}
\begin{document}

\twocolumn[
  \icmltitle{Canzona: A Unified, Asynchronous, and Load-Balanced Framework for Distributed Matrix-based Optimizers}

  \icmlsetsymbol{equal}{*}

  \begin{icmlauthorlist}
    \icmlauthor{Liangyu Wang}{equal,kaust}
    \icmlauthor{Siqi Zhang}{equal,ali}
    \icmlauthor{Junjie Wang}{pku}
    \icmlauthor{Yiming Dong}{ali}
    \icmlauthor{Bo Zheng}{ali}
    \icmlauthor{Zihan Qiu}{ali}
    \icmlauthor{Shengkun Tang}{mbzuai}
    \icmlauthor{Di Wang}{kaust}
    \icmlauthor{Rui Men}{ali}
    \icmlauthor{Dayiheng Liu}{ali}
  \end{icmlauthorlist}

  \icmlaffiliation{kaust}{KAUST}
  \icmlaffiliation{ali}{Alibaba Group}
  \icmlaffiliation{pku}{Peking University}
  \icmlaffiliation{mbzuai}{MBZUAI}

  \icmlcorrespondingauthor{Dayiheng Liu}{liudayiheng.ldyh@alibaba-inc.com}
  \icmlcorrespondingauthor{Rui Men}{menrui.mr@alibaba-inc.com}

  \vskip 0.2in
]

\printAffiliationsAndNotice{}
\let\thefootnote\relax\footnotetext{Work done when Liangyu Wang, Junjie Wang, and Shengkun Tang were interns at Alibaba Group.}

\begin{abstract}
The scaling of Large Language Models (LLMs) drives interest in matrix-based optimizers (e.g., Shampoo, Muon, SOAP) for their convergence efficiency; yet their requirement for holistic updates conflicts with the tensor fragmentation in distributed frameworks like Megatron.
Existing solutions are suboptimal: synchronous approaches suffer from computational redundancy, while layer-wise partitioning fails to reconcile this conflict without violating the geometric constraints of efficient communication primitives. To bridge this gap, we propose Canzona, a Unified, Asynchronous, and Load-Balanced framework that decouples logical optimizer assignment from physical parameter distribution. For Data Parallelism, we introduce an $\alpha$-Balanced Static Partitioning strategy that respects atomicity while neutralizing the load imbalance. For Tensor Parallelism, we design an Asynchronous Compute pipeline utilizing Micro-Group Scheduling to batch fragmented updates and hide reconstruction overhead. Extensive evaluations on the Qwen3 model family (up to 32B parameters) on 256 GPUs demonstrate that our approach preserves the efficiency of established parallel architectures, achieving a $1.57\times$ speedup in end-to-end iteration time and reducing optimizer step latency by $5.8\times$ compared to the baseline.
\end{abstract}

\section{Introduction}
\label{sec:introduction}

\begin{figure*}[h]
    \centering
    \includegraphics[width=0.99\textwidth]{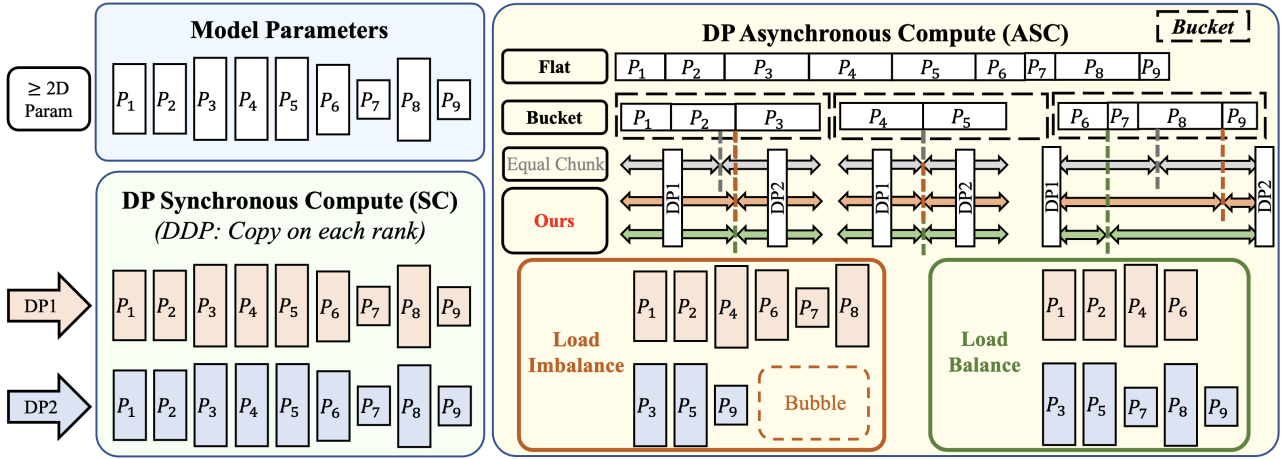}
    \caption{\textbf{Comparison of Data Parallelism (DP) Partitioning Strategies.}
      \textbf{(Left) DP-SC:} One way that can be directly used by those optimizers is DDP, which replicates optimizer states on all ranks, resulting in \textbf{Redundant Compute} where every rank performs identical matrix-based operations (synchronous).
      \textbf{(Right) DP-ASC:} Eliminates redundancy by partitioning states.
      \textbf{Equal Chunk (Standard ZeRO-1, \textcolor{Black!70}{Gray Arrow and Line}):} Standard partitioning (e.g., for \underline{AdamW}) slices the buffer into uniform shards ($|B|/R$). This arbitrary slicing (\textit{\textbf{\textcolor{Black!70}{dashed lines}}}) violates the atomicity required by matrix-based optimizers.
      \textbf{Ours (Static Partitioning, \textcolor{BurntOrange}{Orange} / \textcolor{OliveGreen}{Green} Arrow):} We enforce parameter atomicity by respecting tensor boundaries.
      \textit{Load Imbalance (\textbf{\textcolor{BurntOrange}{Orange Arrow, Line, and Box}}):} A naive atomic assignment leads to significant computational stragglers and communication bubbles (\textit{\textbf{\textcolor{BurntOrange}{dashed box}}}) due to varying parameter costs.
      \textit{Load Balance (\textbf{\textcolor{OliveGreen}{Green Arrow, Line, and Box}}):} Our $\alpha$-Balanced algorithm optimizes the static layout, redistributing whole parameters to equalize the workload across ranks.
      \textbf{Note:} In this figure, the blocks labeled $P$ represent the \textbf{optimizer states} and the associated \textbf{update computation} for those parameters. The parameters themselves remain replicated across ranks during the forward and backward passes (following the ZeRO-1 protocol).}
    \label{fig:async-dp}
\end{figure*}

The relentless scaling of Large Language Models (LLMs)~\citep{qwen2025qwen25technicalreport, yang2025qwen3, grattafiori2024llama, guo2025deepseek, kaplan2020scaling, hoffmann2022training} has driven a continuous search for more efficient training algorithms. While element-wise optimizers like AdamW~\citep{loshchilov2017decoupled} and SGD~\citep{ruder2016overview} remain the de facto standard, recent advances~\citep{liu2025muon, anil2020scalable, osawa2019large, wen2025fantastic} have demonstrated the superior convergence efficiency of \textbf{Matrix-based Optimizers}, such as Shampoo~\citep{gupta2018shampoo}, SOAP~\citep{vyas2024soap}, and Muon~\citep{jordan6muon}. Unlike element-wise optimizers that update parameters independently, these algorithms leverage second-order information or structural properties (e.g., via SVD~\citep{golub2013matrix} or Newton-Schulz iterations~\citep{higham1997stable}) to accelerate training.

However, deploying matrix-based optimizers in modern massive-scale training stacks presents a fundamental system-algorithm conflict. To minimize memory footprints, state-of-the-art frameworks like Megatron~\citep{shoeybi2019megatron} employ aggressive sharding strategies: ZeRO-1~\citep{rajbhandari2020zero, zhao2023pytorch} partitions optimizer states across Data Parallelism (DP) ranks, while Tensor Parallelism (TP) splits weight matrices across devices~\citep{shoeybi2019megatron}. This sharding creates a strict \textbf{Atomicity Constraint}: matrix-based optimizers mathematically require access to the full tensor dimensions to perform holistic updates, but the distributed system physically fragments these tensors across disparate ranks. The details are illustrated in Appendix~\ref{sec:extended-preliminary}.

Existing approaches struggle to reconcile this conflict without severe compromises. Naive strategies often resort to Synchronous Compute (SC) (see details in Section~\ref{subsec:dp-design_analysis}), where ranks perform redundant operations or block globally to preserve atomicity, significantly hindering scalability. Alternatively, solutions like layer-wise partitioning attempt to distribute the workload but fundamentally violate the geometric layout required by ZeRO primitives (see detailed analysis in Section~\ref{subsec:dp-design_analysis} and Appendix~\ref{subsec:layerwise-geometric-conflict}). 
By misaligning the global load-balancing with the physical bucket-based parameter partition, they lose the capability for efficient, bucket-based \texttt{Reduce-Scatter}, sacrificing system throughput for algorithmic correctness.

To bridge this gap, we propose a \textbf{Unified, Asynchronous, and Load-Balanced} framework designed to support generic matrix-based optimizers while preserving the efficiency of established parallel architectures. Our method decouples the logical assignment of optimizer tasks from the physical distribution of parameters. For DP, we introduce a Static Partitioning strategy that respects the \textbf{Atomicity Constraint} by assigning whole parameters to ranks, thereby enabling zero-communication local updates (Figure~\ref{fig:async-dp}, ``Ours''). For TP, we design an Asynchronous Compute pipeline that batches fragmented tensor updates into Micro-Groups, effectively hiding the communication overhead of reconstruction (Figure~\ref{fig:async-tp}, Right). Crucially, our design respects both the Atomicity Constraint and the ZeRO-1 Geometric Constraints (see details in Section~\ref{subsec:dp-design_analysis}), allowing us to fully inherit the efficient, coalesced communication overlapping capabilities of the standard Forward-Backward pass.

A significant challenge introduced by this atomic assignment is load imbalance.
Naive assignment leads to severe stragglers and computational bubbles, as depicted in Section~\ref{subsec:dp-design_analysis} and \ref{subsec:tp-task_abstraction}. To address this, we propose Canzona, a unified, asynchronous, and load-Balanced framework for Distributed matrix-based optimizer. We formulate the partitioning as a load-balancing optimization problem. We propose an $\alpha$-Balanced Greedy LPT~\citep{graham1969bounds} algorithm for DP and a Micro-Group Scheduling algorithm for TP to equalize execution times across heterogeneous workloads.
Our contributions are summarized as follows:
\begin{itemize}
    \item \textbf{A Unified and Decoupled System Architecture:} We propose a unified distributed framework, \textbf{Canzona}, that reconciles the \underline{Atomicity Constraint} of matrix-based optimizers with established ZeRO and TP paradigms. By fundamentally decoupling the logical optimizer assignment from the physical parameter distribution, our \textit{Static Partitioning} (for DP) enables zero-communication during optimizer updates and preserves ZeRO-style geometric alignment between bucketed parameters and optimizer ownership, while our \textit{Asynchronous Micro-Group Pipeline} (for TP) performs communication-efficient optimizer update via fused All-to-All. Crucially, Canzona provides a unified, optimizer-agnostic abstraction that assigns each optimizer task a designated host rank and executes these tasks asynchronously in parallel across ranks, enabling compute–compute overlap that substantially reduces optimizer-step makespan.
    \item \textbf{Heterogeneity-Aware Load Balancing Algorithms:} We identify that the non-linear computational complexity (e.g., cubic scaling) of matrix-based operators introduces severe workload heterogeneity, which naive partitioning fails to handle. We formulate this as a static scheduling problem and propose two dedicated offline algorithms: the \underline{$\alpha$-Balanced Greedy LPT} algorithm for DP, and a \underline{Micro-Group Balanced Scheduling} algorithm with greedy rollback for TP. These strategies effectively neutralize computational stragglers and eliminate pipeline bubbles.
    \item \textbf{Large-Scale Verification and Efficiency:} We evaluate our framework on a cluster of 256 GPUs training the Qwen3 model family (ranging from 1.7B to 32B parameters). Extensive experiments demonstrate that our strategy outperforms \texttt{layerwise\_optimizer}, achieving a \textbf{1.57x} speedup in end-to-end iteration time and reducing the specific optimizer step latency by \textbf{5.8x}. We further validate the generality of our approach across multiple emerging optimizers, including Muon, Shampoo, and SOAP.
\end{itemize}

\section{Preliminary}
\label{sec:preliminary}

Modern LLM training relies on Data Parallelism (DP) to scale training, yet its vanilla implementation, Distributed Data Parallelism (DDP), necessitates redundant state replication (Figure~\ref{fig:async-dp}, Left). 
To improve memory efficiency, frameworks like Megatron adopt the ``ZeRO-1'' strategy, utilizing a contiguous \texttt{param\_and\_grad\_buffer} organized into logical ``Buckets'' to pipeline communication (Reduce-Scatter/All-Gather) with computation. 
Complementing this, Tensor Parallelism (TP) is employed orthogonal to DP to uniformly split weight matrices across devices. 
However, a critical system-algorithm conflict arises: standard sharding strategies, specifically the ``Equal Chunk'' ZeRO-1 partition shown in Figure~\ref{fig:async-dp} and the uniform splitting in TP, arbitrarily fragment tensors to balance memory and computation without regard for parameter boundaries. 
This violation of the \textbf{Atomicity Constraint} precludes the use of \textbf{Matrix-based Optimizers} (e.g., Muon and Shampoo), which mathematically require holistic tensor access for matrix-based operations, thereby preventing efficient local updates. 
We provide a comprehensive review of these distributed training infrastructures and matrix-based optimizations in Appendix~\ref{sec:extended-preliminary}

\section{Load-Balanced Asynchronous Compute for Data Parallelism Matrix-based Optimizer}
\label{sec:dp-method}

To satisfy the atomicity requirement of matrix-based optimizers at scale, we propose a decoupled execution framework that separates logical optimizer-task ownership from physical parameter placement. Under modern parallel training, atomicity-induced communication arises in two planes: (i) the DP plane, where parameters and optimizer states are sharded across data-parallel ranks (e.g., ZeRO-1-style bucketed layouts), and (ii) the TP plane, where individual tensors are partitioned across tensor-parallel ranks and must be reconstructed for holistic matrix updates.

We address these planes with complementary mechanisms. (1) DP (this section): we introduce Static Partitioning that respects both the Atomicity Constraint and the ZeRO-1 geometric bucket layout, so that each matrix update is executed locally by its designated owner rank without introducing additional collectives during the optimizer step, while retaining the training framework’s bucketed communication structure. (2) TP (Section~\ref{sec:tp-method}): we assign each tensor-level optimizer task to a host rank and execute tasks asynchronously and in parallel across ranks (with load-balanced scheduling), enabling scalable distributed matrix computation during the optimizer step.

\subsection{Design Paradigm Analysis: Why Static Layout?}
\label{subsec:dp-design_analysis}

While standard ZeRO-1 (``Equal Chunk'' in Figure~\ref{fig:async-dp}) eliminates redundancy by creating uniform shards ($|B|/R$), this arbitrary slicing violates the \textbf{Atomic Constraint} (Appendix~\ref{sec:extended-preliminary}). As illustrated by the gray dashed lines, update responsibilities for tensors (e.g., $P_2, P_8$) are fragmented across ranks, rendering local holistic operations like SVD impossible without expensive reconstruction. To address this, we must adopt a strategy that respects parameter boundaries. We analyze three paradigms below:

\textbf{Paradigm 1: Synchronous/Redundant Compute (The SC Baseline)}
A straightforward way to preserve atomicity is to revert to a DDP-style approach. This strategy replicates optimizer states across ranks, leading to severe computational redundancy where every device performs identical matrix-based operations (Figure~\ref{fig:async-dp} Left). While mathematically correct, this redundancy severely limits scalability.

\textbf{Paradigm 2: Layerwise Partitioning (Layer-based Schedule).}
A recent proposal, such as NVIDIA's \texttt{layerwise\_optimizer}$^1$\footnote{\hyperlink{https://github.com/NVIDIA/Megatron-LM/pull/2241}{$^1$https://github.com/NVIDIA/Megatron-LM/pull/2241}}, attempts to avoid tensor sharding by assigning optimizer states at the granularity of whole layers. 
However, this approach introduces a fundamental \textbf{Geometric Incompatibility} between the logical task assignment (determined by global load balancing) and the physical parameter partition (determined by ZeRO-1 geometry). 
As detailed in Appendix~\ref{subsec:layerwise-geometric-conflict} and visualized in Figure~\ref{fig:geometric-conflict}, if we preserve Megatron’s \texttt{param\_and\_grad\_buffer} and bucket-level coalesced communication, this ZeRO Geometric Incompatibility structurally precludes the use of the efficient, bucket-based \texttt{Reduce-Scatter} pipeline. 
Consequently, it forces the system to fallback to suboptimal communication paths: 
(1) It typically necessitates \texttt{All-Reduce} for gradient synchronization, which incurs $2\times$ the communication volume compared to the \texttt{Reduce-Scatter} used in ZeRO-1 and significantly increases bandwidth requirements; 
and (2) because the updated parameters are not geometrically aligned with their owners, it requires an additional expensive \texttt{All-Gather} or \texttt{Broadcast} during the optimizer step to redistribute weights, introducing further runtime overhead.

\textbf{Paradigm 3: Static Partitioning (Ours).}
To eliminate the aforementioned communication overheads, we adopt a Static Layout strategy corresponding to the \underline{``Ours''} paradigm in Figure~\ref{fig:async-dp}. Instead of ZeRO-1's "equal chunk", we enforce an \textbf{Atomic Ownership Rule} under ZeRO-1 Geometric Constraints: 
Given a "stride" size $S = |B|/R$ (total buffer size divided by ranks), a parameter $p$ is assigned exclusively to rank $r$ based on its starting position in the flattened buffer. Specifically, rank $r$ owns the optimizer states of $p$ if and only if:
\begin{equation}
    (r-1) \cdot S \le \text{Start\_Index}(p) < r \cdot S
\end{equation}
Crucially, by anchoring the ownership directly to the parameter's physical \text{Start\_Index}, this strategy \underline{automatically} satisfies the \textbf{ZeRO-1 Geometric Constraints} (See details in Appendix~\ref{subsec:layerwise-geometric-conflict}). It ensures that every parameter's optimizer states reside fully on a single device (e.g., Rank 1 owns optimizer states of $P_1, P_2$ entirely as shown in Figure~\ref{fig:async-dp}), respecting \textbf{Atomicity Constraint}, without scrambling the sequential rank ordering. This geometric alignment allows the system to fully inherit the efficient, coalesced \texttt{Reduce-Scatter} communication primitive.

Given the prohibitive overheads associated with Paradigms 1 and 2, we adopted the Static-Layout  Partitioning strategy as our implementation.

\textbf{The Challenge: Load Imbalance.}
While preserving atomicity is necessary, a naive static assignment leads to the \textbf{``Load Imbalance''} scenario depicted in Figure~\ref{fig:async-dp}. Because matrix-based optimizers have non-linear computational costs, a rank assigned a heavy parameter (e.g., the large block on Rank 1 in the figure) becomes a computational straggler.
This creates \underline{pipeline bubbles} (highlighted by the dashed box in Figure~\ref{fig:async-dp}) on other ranks, forcing them to wait and severely degrading global throughput, and introduce higher peak memory usage. (We also note that the variable chunk sizes of each bucket (labeled as ``Bucket'' in Figure~\ref{fig:async-dp}) of the \texttt{param\_and\_grad\_buffer} introduce \texttt{Reduce-Scatter} and \texttt{All-Gather} communication imbalance, but this is a secondary concern as it can be effectively hidden by overlapping communication with forward and backward computation. (Appendix~\ref{subsec:extended-exp-dp-alpha})) Our goal is to transform this state into the \textbf{``Load Balance''} state shown in the bottom right of Figure~\ref{fig:async-dp}, where parameters are redistributed to equalize the workload. We formulate the problem as a load-balanced optimization problem below.

\subsection{Load-Balance Optimization}
\label{subsec:dp-load-balance-optimization}

Consider a distributed training setting with $R$ ranks and a \texttt{param\_and\_grad\_buffer} chunked into $N$ logical buckets $\mathcal{B} = \{B_1, \dots, B_N\}$. Each bucket $B_i$ consists of an ordered sequence of non-splitable parameters $P_i$.
We assign a function $\mathcal{W}(p)$ (Typically $\mathcal{W}(p) = numel(p)$) to each parameter $p$, representing its primary resource cost (e.g., memory usage, computation, or communication). Note that while we assume linear costs here for simplicity, we provide a generalized explanation for non-linear optimizer costs in Appendix~\ref{sec:non-linear-generalization}.

\begin{algorithm}[htbp]
\caption{$\alpha$-Balanced Greedy LPT Partitioning (\S~\ref{subsec:alg_walkthrough})}
\label{alg:greedy_lpt}
\begin{algorithmic}[1]
\REQUIRE Buckets $\mathcal{B}$, Ranks $R$, Balance Factor $\alpha$
\REQUIRE Load function $\mathcal{W}(\cdot) = numel(p)$
\ENSURE Partition vectors $\{\mathbf{s}_i\}$ for all $i$

\STATE \textbf{Define:} Let $\mathcal{W}^i = \sum_{p \in B_i} \mathcal{W}(p)$ be the total load of bucket $i$.
\STATE \textbf{Define:} Let $\Phi_i(u) = \sum_{p \in B_i[0:u]} \mathcal{W}(p)$ be the cumulative load up to cut point $u$.

\STATE \textbf{Initialize:} Global load vector $\mathbf{L} \leftarrow \mathbf{0}_R$
\STATE \textbf{Target:} $\mu \leftarrow (\sum_{i} \mathcal{W}^i) / R$
\STATE \textbf{Sort:} Virtual reorder $\mathcal{B}$ such that $\mathcal{W}^{\pi(1)} \ge \mathcal{W}^{\pi(2)} \ge \dots \ge \mathcal{W}^{\pi(N)}$ (LPT based on Load, Virtual Inter-Bucket reorder)

\FOR{$t = 1$ to $N$}
    \STATE Let current bucket index $k = \pi(t)$
    \STATE {\textbf{Step (1): Calculate Deficits (in Load domain)}}
    \STATE $\mathbf{d} \leftarrow [\max(0, \mu - L_r)]_{r=1}^R$; $D_{\text{total}} \leftarrow \sum_{r} \mathbf{d}$
    
    \STATE {\textbf{Step (2): Define Basis Vectors}}
    \STATE $\mathbf{v}_{\text{even}} \leftarrow [1/R, \dots, 1/R]$
    \STATE $\mathbf{v}_{\text{fill}} \leftarrow \mathbf{d} / D_{\text{total}}$ if $D_{\text{total}} > 0$ else $\mathbf{v}_{\text{even}}$
    
    \STATE {\textbf{Step (3): Blended Target Allocation}}
    \STATE $\mathbf{v}^* \leftarrow (1 - \alpha)\mathbf{v}_{\text{even}} + \alpha \mathbf{v}_{\text{fill}}$
    \STATE $\mathbf{target\_alloc} \leftarrow \mathcal{W}^k \cdot \mathbf{v}^*$ \COMMENT{\textit{Target load for ranks}}
    
    \STATE {\textbf{Step (4): Discretization (Project Load to Valid Cuts, following Atomicity Constraint)}}
    \STATE Init $s_{k,0} \leftarrow 0$; $C \leftarrow 0$
    \FOR{$r = 1$ to $R-1$}
        \STATE $C \leftarrow C + \mathbf{target\_alloc}[r]$
        \STATE \COMMENT{\textit{Find cut $u$ where cumulative load $\Phi_k(u)$ is closest to target $C$}}
        \STATE $s_{k,r} \leftarrow \arg\min_{u \in \mathcal{U}_k} | \Phi_k(u) - C |$ \COMMENT{\textit{Update global load with actual load of the slice}}
        \STATE $L_r \leftarrow L_r + (\Phi_k(s_{k,r}) - \Phi_k(s_{k,r-1}))$
    \ENDFOR
    \STATE $s_{k,R} \leftarrow \text{Size}(B_k)$
    \STATE $L_R \leftarrow L_R + (\Phi_k(s_{k,R}) - \Phi_k(s_{k,R-1}))$
\ENDFOR

\STATE \textbf{Return} $\{\mathbf{s}_i\}$ sorted by original indices
\end{algorithmic}
\end{algorithm}

The goal is to determine slicing vectors $\mathbf{s}_i$ for each bucket $B_i$. The cumulative load assigned to rank $r$ for bucket $B_i$ is $L_{i,r} = \sum_{p \in \text{range}(s_{i,r-1}, s_{i,r})} \mathcal{W}(p)$.
We aim to optimize two objectives:

\textbf{(1) Global DP Balance (Minimize Stragglers):} Minimize the maximum deviation from the ideal mean load $\mu_{\text{load}}$:
    \begin{equation}
        \mathcal{J}_{\text{DP}} = \max_{r} \left| \sum_{i=1}^{N} L_{i,r} - \mu_{\text{load}} \right|
    \end{equation}
\textbf{(2) Fwd-Bwd Bucket Communication Balance (Minimize Idle Time):} Ensure data volume within each bucket is distributed evenly:
    \begin{equation}
        \mathcal{J}_{\text{Comm}} = \sum_{i=1}^{N} \sum_{r=1}^{R} \left| S_{i,r} - \frac{|B_i|}{R} \right|
    \end{equation}

Crucially, those optimization objectives respect the ZeRO-1 Geometric Constraints (Appendix~\ref{subsec:layerwise-geometric-conflict}) by \textbf{shifting the slice boundaries} $s_{i,r}$ within the bucket. Because the parameters are never physically reordered, the sequential, monotonic physical ordering is preserved. This "boundary-shifting" mechanism serves as the logical bridge that allows us to neutralize load imbalance while retaining the capability to launch optimal, coalesced \texttt{Reduce-Scatter} primitives.

Optimizing this discrete partitioning problem under strict atomicity constraints is NP-hard. We propose a heuristic approach named $\alpha$-Balanced Greedy LPT (Algorithm~\ref{alg:greedy_lpt}). By processing buckets in Longest Processing Time (LPT) order, we ensure that heavy parameter blocks are handled early when the flexible ``deficits'' space is largest.
The algorithm employs a control parameter $\alpha \in [0, 1]$ to interpolate between two objectives: \textbf{(1) Global Load Balancing ($\alpha \to 1$):} Prioritizes filling the accumulated deficits of straggler ranks to minimize the global makespan and optimizer state memory usage. \textbf{(2) Fwd-Bwd Bucket Communication Balance ($\alpha \to 0$):} Ignores historical accumulation and partitions the current bucket uniformly (approximating standard ZeRO-1 behavior).

\subsection{System Workflow: Static-Layout Enforcement}
\label{subsec:dp-system-workflow}

The execution workflow consists of two stages:

\textbf{1. Offline Planning (Setup Phase).}
Before training, the \textbf{$\alpha$-Balanced Greedy LPT} algorithm computes a Global Partition Map $\Pi$. For each rank $r$ and bucket $B_i$, $\Pi$ defines the memory interval $[s_{i,r-1}, s_{i,r})$ owned by that rank. Crucially, because we enforce the Atomicity Constraint, these intervals strictly align with parameter boundaries.
Based on $\Pi$, we override the standard shard registration mechanism. Instead of allocating uniform chunks, each rank $r$ allocates memory strictly proportional to its assigned load volume $S_{i,r}$, ensuring the physical buffer layout matches the logical partition.

\textbf{2. Runtime Execution with Overlapping.}
During training, the data flow respects the static layout while preserving Megatron's efficient communication overlapping: \textbf{(1) Backward (Variable-Size Reduce-Scatter):} As gradients accumulate in bucket $B_i$, we trigger a non-uniform \texttt{Reduce-Scatter} operation. Unlike the standard implementation, which assumes equal split sizes, our communication primitive handles the variable shard sizes $S_{i,r}$ defined by $\Pi$. Importantly, since Megatron overlaps this communication with the backward computation of the subsequent bucket ($B_{i-1}$), the slight latency variance caused by the non-uniform data sizes is effectively hidden, maintaining high training throughput.
\textbf{(2) Optimizer Step (Load-Balanced Asynchronous Computation and Zero-Communication):} Rank $r$ updates the parameters strictly within its assigned interval. Since the partition respects atomic boundaries, every parameter $p$ (with its gradient and optimizer state) is fully available locally. The matrix-based optimizer performs operations without any additional communication.
\textbf{(3) Forward (Variable-Size All-Gather):} Before the forward pass, a non-uniform \texttt{All-Gather} reconstructs the full bucket parameters. Similar to the backward pass, this operation is overlapped with the forward computation of the previous bucket, minimizing the impact of size imbalance.

\begin{figure*}[h]
    \centering
    \includegraphics[width=0.95\textwidth]{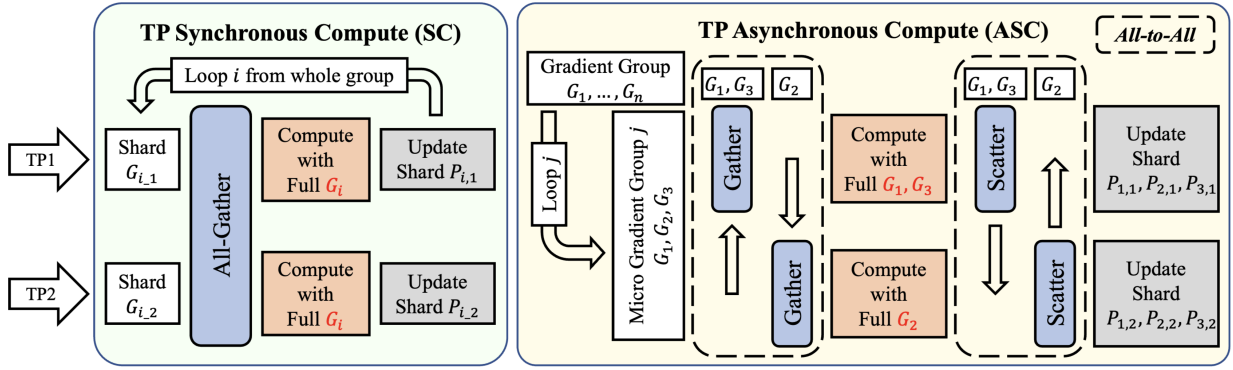}
    \caption{\textbf{Optimizer Update Workflow Comparison of Tensor Parallelism (TP) Strategies.}
      \textbf{(Left) TP-SC:} An intuitive approach that relies on synchronous collective communication (All-Gather) and redundant computation (performing the same tensor operations on all TP ranks), which limits scalability and efficiency.
      \textbf{(Right) TP-ASC:} Our proposed strategy utilizing \textbf{Micro-Group Scheduling}.
      \textbf{Micro Gradient Group:} Gradients are aggregated into micro groups (labeled as $G$ in the figure) to saturate the \textbf{All-to-All} (dashed box) communication bandwidth, replacing the inefficient small-kernel calls.
      \textbf{Load Balancing:} Instead of fixed assignments, these groups are dynamically scheduled to Host Ranks. The distinct block lengths in the "Compute" phase illustrate our algorithm's ability to handle varying computational costs, minimizing the overall execution makespan.}
    \label{fig:async-tp}
\end{figure*}

\section{Tensor Parallelism with Load-Balanced Asynchronous Compute}
\label{sec:tp-method}

While our DP strategy (Section~\ref{sec:dp-method}) strictly enforces a zero-communication layout during optimizer step to protect the global network, Intra-node Tensor Parallelism (TP) presents a different challenge. 
Figure~\ref{fig:async-tp} compares the workflow of standard TP and our proposed strategy. 
In the standard \textbf{TP Synchronous Compute (TP-SC)} (Figure~\ref{fig:async-tp} Left, NVIDIA's \texttt{layerwise\_optimizer} implementation), every rank must communicate and compute the full tensor for every parameter. 
This results in the \textit{Redundant Compute} highlighted in the figure, where all devices perform identical operations (e.g., ``Compute with Full $G_1$'') regardless of the tensor ownership.

In contrast, our \textbf{TP Asynchronous Compute (TP-ASC)} (Figure~\ref{fig:async-tp} Right) eliminates this redundancy through a \underline{statically planned} asynchronous computation pipeline. 
As shown in the ``Micro Gradient Group $j$'' block, we batch multiple tensors (e.g., $G_1, G_2, G_3$) into a single fused communication group. 
Crucially, our offline algorithm (Algorithm~\ref{alg:tp-balance}) assigns these tensors to specific Host Ranks to equalize the load $\mathcal{W}$. 
We emphasize that this reconstruction communication is viable specifically because TP typically operates within the high-bandwidth \textit{Intra-node} domain (e.g., via NVLink). This stands in sharp contrast to DP, which spans bandwidth-constrained \textit{Inter-node} connections, thereby necessitating the strict zero-communication strategy we enforced in Section~\ref{sec:dp-method}.

\subsection{Task Abstraction: The Asynchronous Compute Unit}
\label{subsec:tp-task_abstraction}

To enable fine-grained scheduling, we abstract the update of each TP-split parameter $p$ as an atomic ``Compute Task'' assigned to a specific Host Rank $r$. Since the scheduling decision is static (determined during the initialization phase), we can enforce strict locality for optimizer states. \textbf{Specifically, the full optimizer states (e.g., momentum) will be initialized directly on their designated Host Ranks.} Consequently, these states never require transmission throughout the training process.

\begin{algorithm}[h]
   \caption{Micro-Group Construction for TP load balance with Greedy Rollback (See \S~\ref{subsec:alg_walkthrough} for a detailed walkthrough)}
   \label{alg:tp-balance}
\begin{algorithmic}[1]
   \STATE {\bfseries Input:} Parameters $\mathcal{P}$, Cost Model $\mathcal{W}(\cdot)$, Ranks $R$, Cap $C_{max}$
   \STATE {\bfseries Output:} Sequence of Micro Groups $\mathbb{M}$
   
   \STATE \COMMENT{\textit{Phase 1: Deterministic Global LPT Sort}}
   \STATE $\mathcal{P}_{sorted} \leftarrow \text{Sort}(\mathcal{P}, \text{key}=(\mathcal{W}(p), \text{ID}_p), \text{descending})$
   
   \STATE $M_{curr} \leftarrow \emptyset$; \quad $\mathbb{M} \leftarrow \emptyset$
   
   \STATE \COMMENT{\textit{Phase 2: Partitioning with Rollback}}
   \FOR{$p$ {\bfseries in} $\mathcal{P}_{sorted}$}
       \STATE $M_{test} \leftarrow M_{curr} \cup \{p\}$
       
       \STATE \COMMENT{\textit{Min-Heap Solver to find optimal Host Ranks}}
       \STATE $L_{max} \leftarrow \text{MinHeapBalance}(M_{test}, R)$
       
       \IF{$L_{max} \le C_{max}$}
           \STATE $M_{curr} \leftarrow M_{test}$ \COMMENT{Valid: Accept $p$}
       \ELSE
           \STATE \COMMENT{\textit{Constraint Violated: Rollback \& Finalize}}
           \STATE $\mathbb{M}.\text{append}(M_{curr})$
           \STATE $M_{curr} \leftarrow \{p\}$ \COMMENT{Start new micro group with $p$}
       \ENDIF
   \ENDFOR
   \STATE $\mathbb{M}.\text{append}(M_{curr})$
   \STATE \textbf{return} $\mathbb{M}$
\end{algorithmic}
\end{algorithm}

Combined with the property that optimizer steps derive updates solely from gradients and states without requiring the weight parameters ($P$) immediately, we can optimize the data flow to strictly minimal communication.
The lifecycle of a task group follows the execution stages visualized in the ``Micro Gradient Group $j$'' panel of Figure~\ref{fig:async-tp}:

\textbf{(1) All-to-All for gathering:} Since full optimizer states are already resident on Host Ranks, we only aggregate the \textbf{gradients} $G$. We fuse the asynchronous gather operations of multiple tensors (e.g., Micro Group $j=\{G_1, G_2, G_3\}$ in Figure~\ref{fig:async-tp}) into a single All-to-All collective. By batching parameters into Micro Groups, we ensure the data volume is sufficient to saturate the All-to-All bandwidth, avoiding the overhead of small messages while preventing the prohibitive memory peak caused by fusing all gradients globally. This efficiently routes the gradient shards to their designated Host Ranks where the corresponding optimizer states await.
\textbf{(2) Asynchronous Computation:} Host Ranks execute the matrix-based operations using the gathered gradients and the locally resident optimizer states (e.g., $\Delta W = \text{Muon}_\text{step}(G, \text{States}_{local})$). This phase balances the workload by assigning heavy computations (like $G_2$ in Figure~\ref{fig:async-tp}) to dedicated ranks, while packing lighter tasks ($G_1, G_3$) together.
\textbf{(3) All-to-All for scattering:} Once the update tensors ($\Delta W$) are computed, the Host Ranks slice them into shards. These update shards are then scattered back to the original owners of the parameters via a fused All-to-All operation.
\textbf{(4) Local Update:} Finally, as shown in the ``Update Shard'' block of Figure~\ref{fig:async-tp}, each rank applies the received update shards to its local parameter shards (e.g., $P_{i, local} \leftarrow P_{i, local} + \Delta W_{i, local}$). This guarantees mathematical correctness while avoiding the transmission of both model weights and optimizer states.

\subsection{Workload Scheduling: Hierarchical Partitioning}
\label{subsec:tp-balance}

While Figure~\ref{fig:async-tp} illustrates the example ideal execution flow for a single instance (Micro Group $j$), the system-level challenge lies in generalizing this balance across the entire model. 
We must rigorously partition the full parameter set $\mathcal{P}$ into a sequence of such Micro Groups $\mathbb{M} = \{M_1, \dots, M_K\}$ and determine the optimal Host Rank assignments to minimize the overall execution time. 
We formulate this as a hierarchical partitioning problem with \textbf{Lexicographical Objectives}:

\textbf{Priority 1: Minimize Computational Imbalance.} Within each micro group $M_k$, we minimize the makespan across the $R$ ranks. Let $\mathcal{L}_{k,r}$ be the load on rank $r$ for group $M_k$:
\begin{equation}
    \min \Phi_1(M_k) = \max_{r} (\mathcal{L}_{k,r}) - \min_{r} (\mathcal{L}_{k,r}) \nonumber
\end{equation}
\textbf{Priority 2: Maximize Group Saturation.} To reduce the total number of communication steps, we pack as much load as possible into each micro group without violating the memory/buffer capacity $C_{max}$:
\begin{equation}
    \max \Phi_2(M_k) = \sum_{r=1}^{R} \mathcal{L}_{k,r}, \quad \text{s.t. } \max_{r}(\mathcal{L}_{k,r}) \le C_{max} \nonumber
\end{equation}

\textbf{Algorithm: Global LPT with Greedy Rollback.}
We propose a deterministic heuristic (Algorithm~\ref{alg:tp-balance}) to solve this optimization problem offline. 
First, we use a \textbf{Global Sort} based on computational cost (LPT rule) to ensure determinism across ranks and prioritize heavy tasks (which are harder to balance).
Second, we employ a \textbf{Greedy Packing with Rollback} strategy. We iteratively add parameters to a current micro group and simulate load distribution using a Min-Heap. 
If adding a parameter violates the $C_{max}$ constraint (indicating the makespan or memory has exceeded limits), we trigger a \textit{Rollback}: the current group is finalized, and the parameter initiates a new group.

Note that, similar to Section~\ref{subsec:dp-load-balance-optimization}, while we assume a simplified linear cost model $\mathcal{W}(p) = \text{numel}(p)$ in this section for clarity, our framework formulation fully supports the generalized non-linear complexity (detailed in Appendix~\ref{sec:non-linear-generalization}) inherent to optimizers like Muon and Shampoo.

\textbf{Integration with Runtime Workflow.}
The sequence $\mathbb{M}$ generated by Algorithm~\ref{alg:tp-balance} serves as the static execution plan. 
Effectively, this algorithm automatically chunks the continuous stream of gradients into discrete, load-balanced units---one of which corresponds to the example ``Micro Gradient Group $j$'' visualized in Figure~\ref{fig:async-tp}.
During training, the runtime system simply iterates through this pre-computed sequence $\mathbb{M}$. 
For each group $M_k$, it triggers the exact \textit{Asynchronous Compute Unit} lifecycle described in Section~\ref{subsec:tp-task_abstraction}, ensuring that the complex load balancing decisions made offline are executed efficiently with minimal runtime overhead.

\subsection{Canzona: Framework for Unified, Asynchronous, and Load-Balanced Matrix-based Optimizer}
\label{subsec:method_summary}

We conclude our methodology by highlighting the cohesive design philosophy that spans both parallelism dimensions.
Our framework establishes a strictly \underline{Unified} architecture: neither the DP partitioner (Section~\ref{subsec:dp-load-balance-optimization}) nor the TP scheduler (Section~\ref{subsec:tp-balance}) requires modification to the optimizer's internal mathematics.
Instead, we treat tensor updates as generic computational tasks defined solely by their cost metrics, making the system \underline{optimizer-agnostic} and compatible with any current or future matrix-based optimizers.
Furthermore, both DP and TP strategies converge on a fully \underline{Asynchronous} execution model, ensuring that expensive TensorOps are asynchronously executed.
Finally, by enforcing \underline{Load-Balanced} static planning at both DP and TP levels, we effectively neutralize the cost loads that previously hindered the scaling of matrix-based optimizers.

\begin{figure*}[h]
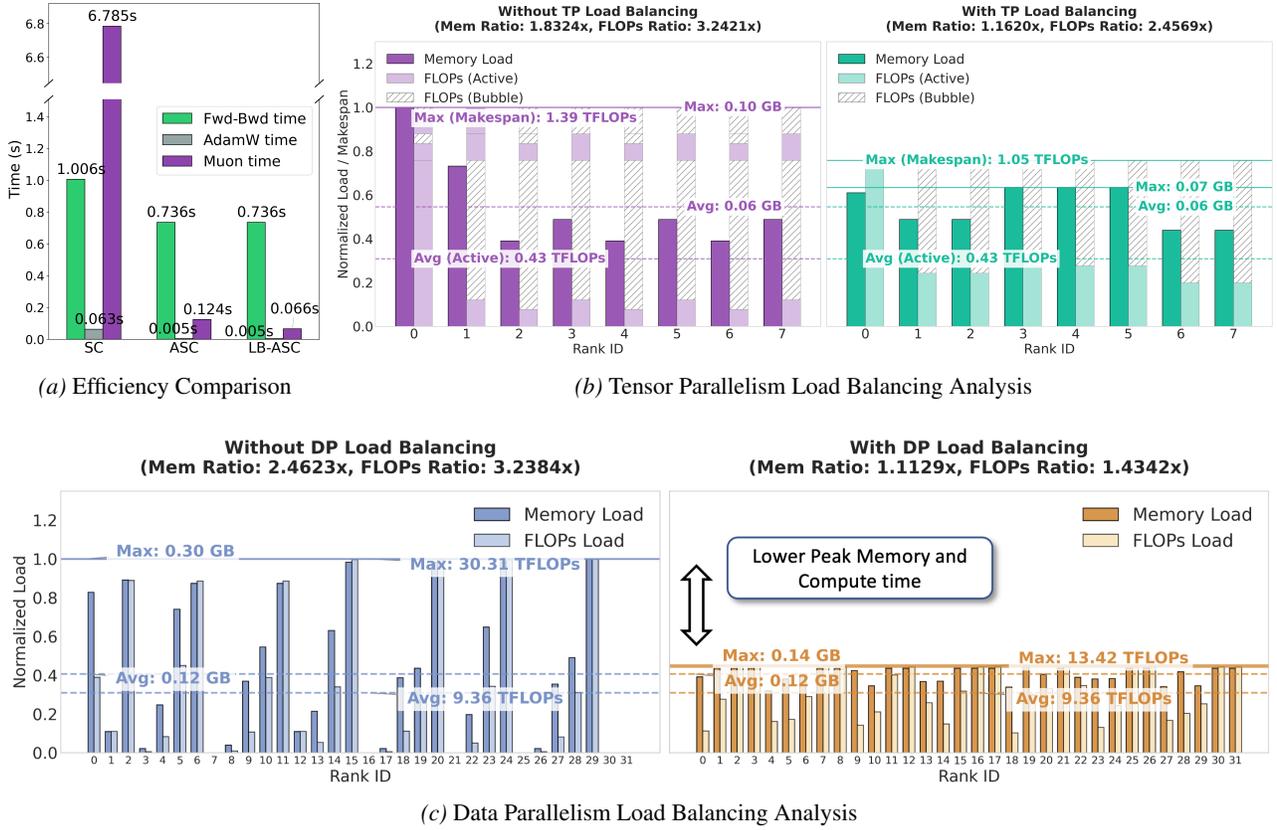

    \centering
    \begin{subfigure}[b]{0.25\textwidth}
        \centering
        \includegraphics[width=\linewidth]{figures/main-exp-efficiency.pdf}
        \caption{Efficiency Comparison}
    \end{subfigure}
    \begin{subfigure}[b]{0.73\textwidth}
        \centering
        \includegraphics[width=\linewidth]{figures/main-exp-tp-lb.pdf}
        \caption{Tensor Parallelism Load Balancing Analysis}
    \end{subfigure}
    
    \par\bigskip
    
    \begin{subfigure}[b]{0.98\textwidth}
        \centering
        \includegraphics[width=\linewidth]{figures/main-exp-dp-lb.pdf}
        \caption{Data Parallelism Load Balancing Analysis}
    \end{subfigure}
    
    \caption{\textbf{Main Results:} \textbf{(a) Efficiency Comparison:} Our LB-ASC strategy outperforms baselines by effectively eliminating computational bubbles and maximizing device utilization.
    \textbf{(b) \& (c) Load Balancing Analysis:} The visualized load distributions demonstrate that our proposed scheduling algorithms successfully flatten the workload variance for both Tensor Parallelism (b) and Data Parallelism (c), significantly mitigating the straggler problem compared to naive partitioning.}
    \label{fig:main_experiments}
\end{figure*}

\section{Experiment}

We evaluate our framework on a cluster with up to 512 GPUs, covering experimental setup (Section~\ref{subsec:exp-setup} and Appendix~\ref{subsec:extended-exp-setup}), efficiency comparison (Section~\ref{subsec:exp-main-results}), and precision verification (Section~\ref{subsec:exp-muon-precision-verify}). We provide extended analyses in \textbf{Appendix~\ref{sec:extended-exp}}: \textbf{(1) Full Performance Comparison} against NVIDIA's \texttt{layerwise\_optimizer} (Appendix~\ref{subsec:extended-exp-compare-nv}); \textbf{(2) Scalability Analysis} across DP/TP and model sizes (Appendix~\ref{subsec:extended-exp-scaling-analysis}); \textbf{(3) Generalization} to Shampoo and SOAP (Appendix~\ref{subsec:extended-exp-generalization-to-other-optim}); and \textbf{(4) Ablation Studies} on hyper-parameters (Appendix~\ref{subsec:extended-exp-dp-alpha}\&\ref{subsec:extended-exp-tp-fuse-size}).

\subsection{Experiment Setup}
\label{subsec:exp-setup}

We evaluate the Qwen3 family (1.7B--32B) using the Muon optimizer (sequence length = 4096, batch size per-DP-rank=1) by default, with verification on Shampoo and SOAP. 
The number of GPUs varies by experiment and is specified in each subsection.
We evaluate \textbf{Canzona} via its core strategy, \textbf{LB-ASC} (Load-Balanced Asynchronous Compute), against:
(1) \textbf{SC} (Synchronous Compute (DDP/TP)); 
(2) \textbf{NV-layerwise} (NVIDIA's \texttt{layerwise\_optimizer}); 
and (3) our ablation \textbf{ASC} (Asynchronous Compute (ZeRO-1/TP) (Decoupled architecture without load balancing). 

\subsection{Main Results}
\label{subsec:exp-main-results}

All experiments in this section are conducted on a large-scale cluster of 256 GPUs. 
The distributed topology is configured with a DP size of 32 and a TP size of 8. 

\textbf{Effectiveness of Load Balancing Strategies (Figure~\ref{fig:main_experiments}).}
We analyze how our proposed strategies mitigate the severe load imbalance inherent in matrix-based optimizers. 
For Data Parallelism (Figure~\ref{fig:main_experiments}c), standard static partitioning (``Without DP Load Balancing'') results in extreme stragglers, exhibiting a FLOPs imbalance ratio (Max/Avg) of \textbf{3.24$\times$} and a Memory imbalance ratio of \textbf{2.46$\times$}. 
In contrast, our $\alpha$-Balanced Partitioning drastically reduces these ratios to \textbf{1.43$\times$} and \textbf{1.11$\times$}, effectively flattening the workload distribution.
For Tensor Parallelism (Figure~\ref{fig:main_experiments}b), the naive baseline similarly suffers from high variance (FLOPs Ratio: 3.24$\times$). 
By introducing our Micro-Group Scheduling, we improve the FLOPs ratio to \textbf{2.46$\times$} and the Memory ratio to \textbf{1.16$\times$}. 
Crucially, as shown in the Makespan comparison (Figure~\ref{fig:main_experiments}a), this improved balance directly translates to efficiency: LB-ASC achieves the lowest maximum step time (1.05 TFLOPS equivalent) compared to SC and ASC, effectively eliminating the computation bubbles.

\begin{figure}[h]
    \centering
    \includegraphics[width=0.6\linewidth]{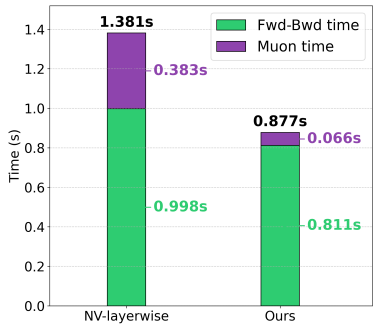}
    \caption{\textbf{End-to-End Iteration Time Comparison:} Our framework significantly outperforms \texttt{layerwise\_optimizer}. 
    The performance advantage is driven by two factors: (1) the elimination of runtime communication during the optimizer step via our decoupled design, and (2) the preservation of the ZeRO-1 Geometric Constraint mentioned in Appendix~\ref{subsec:layerwise-geometric-conflict}.
    }
    \label{fig:comparison_nv}
\end{figure}

\textbf{Comparison with \texttt{layerwise\_optimizer} (Figure~\ref{fig:comparison_nv}).}
We compare the end-to-end training iteration time against NVIDIA's \texttt{layerwise\_optimizer} (``NV-layerwise''), the current best practice for preserving atomicity.
As shown in Figure~\ref{fig:comparison_nv}, our approach achieves a \textbf{1.57$\times$ speedup} in total iteration time (0.877s vs. 1.381s). 
This performance gain stems from two critical sources:
(1) \textbf{Drastic Reduction in Optimizer Latency (5.8$\times$ Speedup):} Our optimizer step time drops from \textbf{0.383s} to \textbf{0.066s}. While NV-layerwise necessitates extra runtime \texttt{Broadcast} or \texttt{All-Gather} to redistribute updated parameters, our zero-communication DP strategy and asynchronous TP pipeline effectively hide the optimization overhead.
(2) \textbf{Improved Fwd-Bwd Efficiency (1.23$\times$ Speedup):} Our method also reduces the Forward-Backward pass time from \textbf{0.998s} to \textbf{0.811s}. This highlights a structural advantage: NV-layerwise suffers from the geometric conflict described in Appendix~\ref{subsec:layerwise-geometric-conflict}, which structurally forces it to rely on gradient All-Reduce (incurring $2\times$ communication volume), whereas our approach utilizes the more efficient, bucket-overlapped \texttt{Reduce-Scatter}.

\subsection{Precision Verification}
\label{subsec:exp-muon-precision-verify}

\begin{figure}[h]
    \centering
    \includegraphics[width=0.8\linewidth]{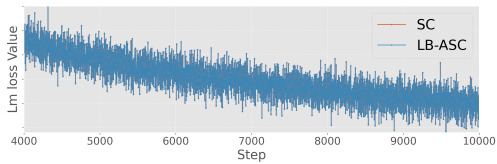}
    \caption{\textbf{Precision Verification:} SC Baseline \& LB-ASC (Ours)}
    \label{fig:precision-muon-qwen3-1.7b}
\end{figure}

To ensure algorithmic correctness, we trained a Qwen3-1.7B model on 400B tokens using the Muon optimizer (DP=8, TP=4). As shown in Figure~\ref{fig:precision-muon-qwen3-1.7b}, the training loss trajectory of our strategy LB-ASC is indistinguishable from the standard synchronous baseline (SC). This confirms that our decoupled partitioning and scheduling strategies act as purely system-level optimizations, preserving the exact convergence behavior of the original matrix-based optimizer. More precision verification experiments can be found in Appendix~\ref{subsec:extended-exp-generalization-to-other-optim}

\section{Related Work}
\label{sec:related_work}

The deployment of matrix-based optimizers in distributed settings presents a unique system-algorithm conflict. While early works like Distributed K-FAC~\citep{osawa2019large} utilized approximate curvature to mitigate communication costs, modern exact optimizers (e.g., Muon, Shampoo) require strict atomicity that conflicts with standard sharding strategies like ZeRO~\citep{rajbhandari2020zero}.
To resolve atomicity, approaches like NVIDIA's \texttt{layerwise\_optimizer} assign states by layer but disregard ZeRO geometric constraints. As detailed in Appendix~\ref{subsec:layerwise-geometric-conflict}, this \textit{Geometric Incompatibility} precludes efficient bucket-based Reduce-Scatter, forcing a fallback to costly All-Reduce ($2\times$ volume). Conversely, our framework harmonizes atomicity with geometric partitioning, fully preserving the optimal communication primitives.
Please refer to Appendix~\ref{sec:extended-related-work} for a comprehensive review.

\section{Conclusion}
\label{sec:conclusion}

In this paper, we presented a unified distributed optimization framework that resolves the fundamental conflict between matrix-based optimizers and modern parallel training strategies. 
By introducing the $\alpha$-Balanced Static Partitioning for DP and the Asynchronous Micro-Group Scheduling for TP, we successfully reconciled strict atomicity requirements with massive parallelism.
Our extensive experiments on the Qwen3 model family demonstrate that our approach not only eliminates the heavy communication overhead of existing solutions but also effectively neutralizes the load imbalance inherent in second-order optimization, achieving a 1.57$\times$ end-to-end speedup.

\nocite{langley00}

\bibliography{main}

\begin{thebibliography}{33}
\providecommand{\natexlab}[1]{#1}
\providecommand{\url}[1]{\texttt{#1}}
\expandafter\ifx\csname urlstyle\endcsname\relax
  \providecommand{\doi}[1]{doi: #1}\else
  \providecommand{\doi}{doi: \begingroup \urlstyle{rm}\Url}\fi

\bibitem[Ahn \& Xu(2025)Ahn and Xu]{ahn2025dion}
Ahn, K. and Xu, B.
\newblock Dion: A communication-efficient optimizer for large models.
\newblock \emph{arXiv e-prints}, pp.\  arXiv--2504, 2025.

\bibitem[Anil et~al.(2020)Anil, Gupta, Koren, Regan, and Singer]{anil2020scalable}
Anil, R., Gupta, V., Koren, T., Regan, K., and Singer, Y.
\newblock Scalable second order optimization for deep learning.
\newblock \emph{arXiv preprint arXiv:2002.09018}, 2020.

\bibitem[Golub \& Van~Loan(2013)Golub and Van~Loan]{golub2013matrix}
Golub, G.~H. and Van~Loan, C.~F.
\newblock \emph{Matrix computations}.
\newblock JHU press, 2013.

\bibitem[Gong et~al.(2025)Gong, Scetbon, Ma, and Meeds]{gong2025towards}
Gong, W., Scetbon, M., Ma, C., and Meeds, E.
\newblock Towards efficient optimizer design for llm via structured fisher approximation with a low-rank extension.
\newblock \emph{arXiv preprint arXiv:2502.07752}, 2025.

\bibitem[Graham(1969)]{graham1969bounds}
Graham, R.~L.
\newblock Bounds on multiprocessing timing anomalies.
\newblock \emph{SIAM journal on Applied Mathematics}, 17\penalty0 (2):\penalty0 416--429, 1969.

\bibitem[Grattafiori et~al.(2024)Grattafiori, Dubey, Jauhri, Pandey, Kadian, Al-Dahle, Letman, Mathur, Schelten, Vaughan, et~al.]{grattafiori2024llama}
Grattafiori, A., Dubey, A., Jauhri, A., Pandey, A., Kadian, A., Al-Dahle, A., Letman, A., Mathur, A., Schelten, A., Vaughan, A., et~al.
\newblock The llama 3 herd of models.
\newblock \emph{arXiv preprint arXiv:2407.21783}, 2024.

\bibitem[Guo et~al.(2025)Guo, Yang, Zhang, Song, Zhang, Xu, Zhu, Ma, Wang, Bi, et~al.]{guo2025deepseek}
Guo, D., Yang, D., Zhang, H., Song, J., Zhang, R., Xu, R., Zhu, Q., Ma, S., Wang, P., Bi, X., et~al.
\newblock Deepseek-r1: Incentivizing reasoning capability in llms via reinforcement learning.
\newblock \emph{arXiv preprint arXiv:2501.12948}, 2025.

\bibitem[Gupta et~al.(2018)Gupta, Koren, and Singer]{gupta2018shampoo}
Gupta, V., Koren, T., and Singer, Y.
\newblock Shampoo: Preconditioned stochastic tensor optimization.
\newblock In \emph{International Conference on Machine Learning}, pp.\  1842--1850. PMLR, 2018.

\bibitem[He et~al.(2025)He, Han, Zhou, Chen, Liu, Chen, and Wang]{he2025root}
He, W., Han, K., Zhou, H., Chen, H., Liu, Z., Chen, X., and Wang, Y.
\newblock Root: Robust orthogonalized optimizer for neural network training.
\newblock \emph{arXiv preprint arXiv:2511.20626}, 2025.

\bibitem[Higham(1997)]{higham1997stable}
Higham, N.~J.
\newblock Stable iterations for the matrix square root.
\newblock \emph{Numerical Algorithms}, 15\penalty0 (2):\penalty0 227--242, 1997.

\bibitem[Hoffmann et~al.(2022)Hoffmann, Borgeaud, Mensch, Buchatskaya, Cai, Rutherford, Casas, Hendricks, Welbl, Clark, et~al.]{hoffmann2022training}
Hoffmann, J., Borgeaud, S., Mensch, A., Buchatskaya, E., Cai, T., Rutherford, E., Casas, D. d.~L., Hendricks, L.~A., Welbl, J., Clark, A., et~al.
\newblock Training compute-optimal large language models.
\newblock \emph{arXiv preprint arXiv:2203.15556}, 2022.

\bibitem[Jordan et~al.()Jordan, Jin, Boza, Jiacheng, Cecista, Newhouse, and Bernstein]{jordan6muon}
Jordan, K., Jin, Y., Boza, V., Jiacheng, Y., Cecista, F., Newhouse, L., and Bernstein, J.
\newblock Muon: An optimizer for hidden layers in neural networks, 2024.
\newblock \emph{URL https://kellerjordan. github. io/posts/muon}, 6.

\bibitem[Kaplan et~al.(2020)Kaplan, McCandlish, Henighan, Brown, Chess, Child, Gray, Radford, Wu, and Amodei]{kaplan2020scaling}
Kaplan, J., McCandlish, S., Henighan, T., Brown, T.~B., Chess, B., Child, R., Gray, S., Radford, A., Wu, J., and Amodei, D.
\newblock Scaling laws for neural language models.
\newblock \emph{arXiv preprint arXiv:2001.08361}, 2020.

\bibitem[Khaled et~al.(2025)Khaled, Ozkara, Yu, Hong, and Park]{khaled2025muonbp}
Khaled, A., Ozkara, K., Yu, T., Hong, M., and Park, Y.
\newblock Muonbp: Faster muon via block-periodic orthogonalization.
\newblock \emph{arXiv preprint arXiv:2510.16981}, 2025.

\bibitem[Li et~al.(2020)Li, Zhao, Varma, Salpekar, Noordhuis, Li, Paszke, Smith, Vaughan, Damania, et~al.]{li2020pytorch}
Li, S., Zhao, Y., Varma, R., Salpekar, O., Noordhuis, P., Li, T., Paszke, A., Smith, J., Vaughan, B., Damania, P., et~al.
\newblock Pytorch distributed: Experiences on accelerating data parallel training.
\newblock \emph{arXiv preprint arXiv:2006.15704}, 2020.

\bibitem[Li(2017)]{li2017preconditioned}
Li, X.-L.
\newblock Preconditioned stochastic gradient descent.
\newblock \emph{IEEE transactions on neural networks and learning systems}, 29\penalty0 (5):\penalty0 1454--1466, 2017.

\bibitem[Liu et~al.(2023)Liu, Li, Hall, Liang, and Ma]{liu2023sophia}
Liu, H., Li, Z., Hall, D., Liang, P., and Ma, T.
\newblock Sophia: A scalable stochastic second-order optimizer for language model pre-training.
\newblock \emph{arXiv preprint arXiv:2305.14342}, 2023.

\bibitem[Liu et~al.(2025)Liu, Su, Yao, Jiang, Lai, Du, Qin, Xu, Lu, Yan, et~al.]{liu2025muon}
Liu, J., Su, J., Yao, X., Jiang, Z., Lai, G., Du, Y., Qin, Y., Xu, W., Lu, E., Yan, J., et~al.
\newblock Muon is scalable for llm training.
\newblock \emph{arXiv preprint arXiv:2502.16982}, 2025.

\bibitem[Loshchilov \& Hutter(2017)Loshchilov and Hutter]{loshchilov2017decoupled}
Loshchilov, I. and Hutter, F.
\newblock Decoupled weight decay regularization.
\newblock \emph{arXiv preprint arXiv:1711.05101}, 2017.

\bibitem[Martens \& Grosse(2015)Martens and Grosse]{martens2015optimizing}
Martens, J. and Grosse, R.
\newblock Optimizing neural networks with kronecker-factored approximate curvature.
\newblock In \emph{International conference on machine learning}, pp.\  2408--2417. PMLR, 2015.

\bibitem[Micikevicius et~al.(2017)Micikevicius, Narang, Alben, Diamos, Elsen, Garcia, Ginsburg, Houston, Kuchaiev, Venkatesh, et~al.]{micikevicius2017mixed}
Micikevicius, P., Narang, S., Alben, J., Diamos, G., Elsen, E., Garcia, D., Ginsburg, B., Houston, M., Kuchaiev, O., Venkatesh, G., et~al.
\newblock Mixed precision training.
\newblock \emph{arXiv preprint arXiv:1710.03740}, 2017.

\bibitem[Osawa et~al.(2019)Osawa, Tsuji, Ueno, Naruse, Yokota, and Matsuoka]{osawa2019large}
Osawa, K., Tsuji, Y., Ueno, Y., Naruse, A., Yokota, R., and Matsuoka, S.
\newblock Large-scale distributed second-order optimization using kronecker-factored approximate curvature for deep convolutional neural networks.
\newblock In \emph{Proceedings of the IEEE/CVF Conference on Computer Vision and Pattern Recognition}, pp.\  12359--12367, 2019.

\bibitem[Qwen et~al.(2025)Qwen, :, Yang, Yang, Zhang, Hui, Zheng, Yu, Li, Liu, Huang, Wei, Lin, Yang, Tu, Zhang, Yang, Yang, Zhou, Lin, Dang, Lu, Bao, Yang, Yu, Li, Xue, Zhang, Zhu, Men, Lin, Li, Tang, Xia, Ren, Ren, Fan, Su, Zhang, Wan, Liu, Cui, Zhang, and Qiu]{qwen2025qwen25technicalreport}
Qwen, :, Yang, A., Yang, B., Zhang, B., Hui, B., Zheng, B., Yu, B., Li, C., Liu, D., Huang, F., Wei, H., Lin, H., Yang, J., Tu, J., Zhang, J., Yang, J., Yang, J., Zhou, J., Lin, J., Dang, K., Lu, K., Bao, K., Yang, K., Yu, L., Li, M., Xue, M., Zhang, P., Zhu, Q., Men, R., Lin, R., Li, T., Tang, T., Xia, T., Ren, X., Ren, X., Fan, Y., Su, Y., Zhang, Y., Wan, Y., Liu, Y., Cui, Z., Zhang, Z., and Qiu, Z.
\newblock Qwen2.5 technical report, 2025.
\newblock URL \url{https://arxiv.org/abs/2412.15115}.

\bibitem[Rajbhandari et~al.(2020)Rajbhandari, Rasley, Ruwase, and He]{rajbhandari2020zero}
Rajbhandari, S., Rasley, J., Ruwase, O., and He, Y.
\newblock Zero: Memory optimizations toward training trillion parameter models.
\newblock In \emph{SC20: International Conference for High Performance Computing, Networking, Storage and Analysis}, pp.\  1--16. IEEE, 2020.

\bibitem[Ruder(2016)]{ruder2016overview}
Ruder, S.
\newblock An overview of gradient descent optimization algorithms.
\newblock \emph{arXiv preprint arXiv:1609.04747}, 2016.

\bibitem[Shi et~al.(2023)Shi, Lee, Iwasaki, Gallego-Posada, Li, Rangadurai, Mudigere, and Rabbat]{shi2023distributed}
Shi, H.-J.~M., Lee, T.-H., Iwasaki, S., Gallego-Posada, J., Li, Z., Rangadurai, K., Mudigere, D., and Rabbat, M.
\newblock A distributed data-parallel pytorch implementation of the distributed shampoo optimizer for training neural networks at-scale.
\newblock \emph{arXiv preprint arXiv:2309.06497}, 2023.

\bibitem[Shoeybi et~al.(2019)Shoeybi, Patwary, Puri, LeGresley, Casper, and Catanzaro]{shoeybi2019megatron}
Shoeybi, M., Patwary, M., Puri, R., LeGresley, P., Casper, J., and Catanzaro, B.
\newblock Megatron-lm: Training multi-billion parameter language models using model parallelism.
\newblock \emph{arXiv preprint arXiv:1909.08053}, 2019.

\bibitem[Vyas et~al.(2024)Vyas, Morwani, Zhao, Kwun, Shapira, Brandfonbrener, Janson, and Kakade]{vyas2024soap}
Vyas, N., Morwani, D., Zhao, R., Kwun, M., Shapira, I., Brandfonbrener, D., Janson, L., and Kakade, S.
\newblock Soap: Improving and stabilizing shampoo using adam.
\newblock \emph{arXiv preprint arXiv:2409.11321}, 2024.

\bibitem[Wang et~al.(2025)Wang, Zhou, Dong, Li, Li, Zhou, Lao, Fang, and Lin]{wang2025conda}
Wang, J., Zhou, P., Dong, Y., Li, H., Li, J., Zhou, X., Lao, Q., Fang, C., and Lin, Z.
\newblock Conda: Column-normalized adam for training large language models faster.
\newblock \emph{arXiv preprint arXiv:2509.24218}, 2025.

\bibitem[Wen et~al.(2025)Wen, Hall, Ma, and Liang]{wen2025fantastic}
Wen, K., Hall, D., Ma, T., and Liang, P.
\newblock Fantastic pretraining optimizers and where to find them.
\newblock \emph{arXiv preprint arXiv:2509.02046}, 2025.

\bibitem[Xie et~al.(2026)Xie, Luo, Tang, Hu, Liu, Ren, Wang, Zhao, Yan, Su, et~al.]{xie2026controlled}
Xie, T., Luo, H., Tang, H., Hu, Y., Liu, J.~K., Ren, Q., Wang, Y., Zhao, W.~X., Yan, R., Su, B., et~al.
\newblock Controlled llm training on spectral sphere.
\newblock \emph{arXiv preprint arXiv:2601.08393}, 2026.

\bibitem[Yang et~al.(2025)Yang, Li, Yang, Zhang, Hui, Zheng, Yu, Gao, Huang, Lv, et~al.]{yang2025qwen3}
Yang, A., Li, A., Yang, B., Zhang, B., Hui, B., Zheng, B., Yu, B., Gao, C., Huang, C., Lv, C., et~al.
\newblock Qwen3 technical report.
\newblock \emph{arXiv preprint arXiv:2505.09388}, 2025.

\bibitem[Zhao et~al.(2023)Zhao, Gu, Varma, Luo, Huang, Xu, Wright, Shojanazeri, Ott, Shleifer, et~al.]{zhao2023pytorch}
Zhao, Y., Gu, A., Varma, R., Luo, L., Huang, C.-C., Xu, M., Wright, L., Shojanazeri, H., Ott, M., Shleifer, S., et~al.
\newblock Pytorch fsdp: experiences on scaling fully sharded data parallel.
\newblock \emph{arXiv preprint arXiv:2304.11277}, 2023.

\end{thebibliography}
\bibliographystyle{icml2026}

\newpage
\appendix
\onecolumn

\section{Summary of Notations}

\begin{table}[htbp]
    \centering
    \caption{Summary of Notations}
    \label{tab:notation}
    \renewcommand{\arraystretch}{1.2}
    \begin{tabular}{l|l}
    \toprule
    \textbf{Symbol} & \textbf{Description} \\
    \midrule
    \multicolumn{2}{c}{\textit{System Settings \& Preliminaries}} \\
    \midrule
    $R$ & Total number of ranks (DP ranks (Section~\ref{sec:dp-method}) or TP ranks (Section~\ref{sec:tp-method})) \\
    $\mathcal{B}$ & The set of logical buckets $\{B_1, B_2, \dots, B_N\}$ in the \texttt{param\_and\_grad\_buffer} \\
    $B_i$ & The $i$-th logical bucket consisting of ordered parameters \\
    $P_i$ & The sequence of non-splitable parameters $(p_{i,1}, \dots, p_{i, M_i})$ in $B_i$ \\
    $S$ & Uniform shard size ($|B|/R$) in standard ZeRO partition \\
    $\mathcal{I}_r$ & The memory interval $[(r-1)S, rS)$ assigned to rank $r$ in ZeRO \\
    $o_p$ & The starting memory offset (index) of parameter $p$ in the flattened buffer \\
    
    \midrule
    \multicolumn{2}{c}{\textit{Load Cost Metrics}} \\
    \midrule
    $\mathcal{C}_{\text{mem}}(p)$ & Memory usage cost of parameter $p$ \\
    $\mathcal{C}_{\text{comp}}(p)$ & Computational cost of parameter $p$ (FLOPs) \\
    $\mathcal{C}_{\text{comm}}(p)$ & Communication cost of parameter $p$ (parameter size, $= numel(p)$) \\
    $\mathcal{W}_{\text{load}}(p)$ & Primary metric for load balancing ($\mathcal{C}_{\text{mem}}$ or $\mathcal{C}_{\text{comp}}$) \\
    $\mathcal{W}_{\text{size}}(p)$ & Secondary metric for communication grouping (usually $\mathcal{C}_{\text{comm}}$) \\
    $\mathcal{W}(p)$ & A general cost metric assigned to parameter $p$ for load balancing \\
    
    \midrule
    \multicolumn{2}{c}{\textit{Problem Formulation \& Load-Balance Algorithm for DP LB-ASC}} \\
    \midrule
    $\mathbf{s}_i$ & Slicing vector $[s_{i,0}, \dots, s_{i,R}]$ containing cut points for bucket $B_i$ \\
    $\mathcal{U}_i$ & Set of feasible (atomic) cut points for bucket $B_i$ \\
    $L_{i,r}$ & The cumulative load assigned to rank $r$ in bucket $B_i$ \\
    $S_{i,r}$ & The data size assigned to rank $r$ in bucket $B_i$. Here, $S_{i,r} = s_{i,r} - s_{i,r-1}$ \\
    $\alpha$ & Control parameter $\in [0, 1]$ trading off between $\mathcal{J}_{\text{DP}}$ and $\mathcal{J}_{\text{Comm}}$ \\
    $\mathbf{L}$ & Vector tracking the global accumulated load for each rank \\
    $\mathbf{v}_{\text{even}}$ & Basis vector for perfectly even distribution \\
    $\mathbf{v}_{\text{fill}}$ & Basis vector for deficit-filling distribution \\
    $\mathbf{v}^*$ & Blended target allocation vector for the current bucket \\
    $\Phi_i(u)$ & Cumulative load function of bucket $B_i$ up to cut point $u$ \\
    \midrule
    \multicolumn{2}{c}{\textit{Problem Formulation \& Load-Balance Algorithm for TP LB-ASC}} \\
    \midrule
    $\mathbb{M}$ & The sequence of Micro Groups $\{M_{1}, ..., M_{K}\}$ generated by the scheduler \\
    $C_{max}$ & The maximum capacity constraint (memory or FLOPs) for a single Micro Group \\
    \bottomrule
    \end{tabular}
\end{table}

\section{Extended Preliminary}
\label{sec:extended-preliminary}

\subsection{Data Parallelism and ZeRO}

Data Parallelism (DP) is the dominant paradigm for training Large Language Models (LLMs). In standard Distributed Data Parallelism (DDP)~\citep{li2020pytorch} (Figure~\ref{fig:async-dp}), the model parameters and optimizer states are replicated across all parallel data ranks. While this enables parallel gradient computation, the redundancy of optimizer states consumes a significant portion of GPU memory—often 2--3$\times$ the model size for mixed-precision training (e.g., FP32 master weights~\citep{micikevicius2017mixed} and momentum).

Zero Redundancy Optimizer (ZeRO-1) \citep{rajbhandari2020zero} addresses this redundancy by partitioning the model states across data-parallel ranks. Specifically, ZeRO-1 (Optimizer State Partitioning) shards the optimizer states such that each rank is responsible for updating only a fraction ($1/R$) of the total parameters. This linear reduction in memory footprint is crucial for scaling typically massive models.

Frameworks like Megatron~\citep{shoeybi2019megatron} implement ZeRO-1 through its \texttt{DistributedOptimizer}, which relies on a unified memory management structure known as the \texttt{param\_and\_grad\_buffer} ("Flat" and "Bucket" boxes in Figure~\ref{fig:async-dp}). (Because ZeRO-1 has become a default choice in Megatron, we will refer to ZeRO-1 as DP in the following text.)

\textbf{Continuous Memory Buffer and Bucketing of \texttt{param\_and\_grad\_buffer}.}
Instead of managing memory for each parameter tensor individually, Megatron flattens all model parameters (and their gradients) into several continuous memory buffers. To facilitate communication overlapping of forward and backward, each continuous buffer is logically divided into $N$ \textit{buckets}. 
As parameters are registered, they are packed into these buckets sequentially. This contiguous layout is essential for high-performance collective communications, allowing a single large operation (e.g., \texttt{Reduce-Scatter}) to process multiple parameters simultaneously rather than launching many small kernels.

\textbf{Uniform Partition Rule and ZeRO-1 Geometric Constraint.} 
Crucially, Megatron imposes a strict geometric partition rule. For a bucket of size $B$, the buffer is rigidly divided into $R$ equal contiguous segments of size $S = |B|/R$, entirely agnostic to the boundaries of individual parameter tensors. Consequently, rank $r$ is hard-wired to process the $r$-th segment. This rigid positional mapping forms a fundamental \textbf{ZeRO-1 Geometric Constraint} that is essential for efficient communication primitives like \texttt{Reduce-Scatter}.

The typical training workflow with this buffer structure proceeds as follows:
\begin{itemize}
    \item \textbf{Backward Pass:} As the backward pass proceeds, gradients are computed and accumulated into the buckets. Once the gradients for all parameters in a bucket are ready, a \texttt{Reduce-Scatter} operation is triggered across all DP ranks. This operation aggregates the gradients and shards them, such that each rank receives only its assigned continuous segment $S$. Importantly, this communication is overlapped with the backward computation of the subsequent bucket.
    \item \textbf{Optimizer Step:} Rank $r$ performs the optimizer step (updating parameters) strictly on its assigned memory range $[(r-1)S, r S)$.
    \item \textbf{Forward Pass:} Before the forward computation for a bucket can begin, the updated parameter shards must be gathered from all ranks. An \texttt{All-Gather} operation is triggered in the first micro-step to reconstruct the full parameters for that bucket. Similar to the backward pass, this communication is overlapped with the forward computation of the previous bucket.
\end{itemize}

\subsection{Tensor Parallelism}

While Data Parallelism distributes the batch dimension, Tensor Parallelism (TP) reduces the memory footprint of activation and parameters by partitioning the model execution itself across multiple devices. Megatron introduces an efficient TP scheme tailored for Transformer architectures, splitting individual weight matrices across multiple GPUs.

Specifically, Megatron utilizes two splitting strategies: \textit{Column Parallelism} and \textit{Row Parallelism}. In a Multi-Head Attention (MHA) or Feed-Forward Network (FFN) block, the first linear layer is typically column-parallelized (splitting the output dimension), while the second linear layer is row-parallelized (splitting the input dimension). This arrangement allows the output of the first layer to be fed directly into the second layer without synchronization, requiring only a single \texttt{All-Reduce} communication operation in the forward pass (and one in the backward pass) per block.
TP is orthogonal to DP; typically, DP is applied across nodes while TP is applied within a node to minimize the latency of the frequent \texttt{All-Reduce} operations.

\subsection{Matrix-based Optimizers}
\label{subsec:tensor_wise_opt}

\textbf{Compatibility with Element-wise Optimizers.}
This rigid partition scheme works seamlessly for standard optimizers like AdamW~\citep{loshchilov2017decoupled} or SGD~\citep{ruder2016overview}. These optimizers are \textit{element-wise}: the update for a parameter scalar $w$ depends solely on its gradient $g$ and its historical states (e.g., momentums $m, v$).
\begin{equation}
    w_{t+1}^{(i)} = \text{Update}(w_t^{(i)}, g_t^{(i)}, m_t^{(i)}, v_t^{(i)})
\end{equation}
Since the operations on the $i$-th element are independent of the $j$-th element, cutting a parameter tensor at an arbitrary index does not affect mathematical correctness. The \texttt{DistributedOptimizer} can essentially treat the entire model as one giant 1D vector without awareness of tensor boundaries.

\textbf{Matrix-based Optimizers.}
Recent advances in optimization have introduced algorithms that move beyond element-wise updates to utilize second-order information or structural properties of the weight matrices. We refer to these as \textbf{Matrix-based Optimizers}.

Examples include Muon~\citep{jordan6muon}, which applies Newton-Schulz iterations to orthogonalize weight matrices, and Shampoo~\citep{gupta2018shampoo}, which utilizes preconditioning matrices derived from the Kronecker product of gradients.

Unlike element-wise, these optimizers operate at \textbf{tensor granularity}. For a parameter $W \in \mathbb{R}^{d_{\text{in}} \times d_{\text{out}}}$, the update rule typically involves MatrixOp like Singular Value Decomposition (SVD) or Matrix Multiplication:
\begin{equation}
    W_{t+1} = W_t - \eta \cdot \text{MatrixOp}(W_t, \nabla W_t)
\end{equation}
This imposes an \textbf{Atomicity Constraint}: the optimizer requires access to the complete dimensions of the tensor $W$ to perform the update. The arbitrary slicing performed by the standard Megatron \texttt{param\_and\_grad\_buffer}—which might cut a single parameter $W$ across two different ranks—renders the local computation of $\text{MatrixOp}$ impossible without expensive on-the-fly communication to reconstruct the tensor. Consequently, adapting DP or TP for matrix-based optimizers requires a partition strategy that respects parameter boundaries.

\section{More Experiments}
\label{sec:extended-exp}

\subsection{Full Experiment Setup}
\label{subsec:extended-exp-setup}

\paragraph{Models and Optimizers.}
We evaluate the scalability and efficiency of our framework using the \textbf{Qwen3}~\citep{yang2025qwen3} model family, scaling from 1.7B to 32B parameters. 
To verify the generality of our approach across different matrix-based algorithms, we conduct experiments with three representative second-order optimizers: \textbf{Muon}~\citep{jordan6muon}, \textbf{Shampoo}~\citep{gupta2018shampoo}, and \textbf{SOAP}~\citep{vyas2024soap}.
Most of the experiments primarily report the detailed performance analysis using \textbf{Qwen3-32B} trained with the \textbf{Muon} optimizer, unless otherwise specified.
The training configuration utilizes a batch-size per DP rank of 1 and a sequence length of 4096.

\paragraph{Baselines and Methods.}
We compare three distinct optimization strategies to demonstrate the effectiveness of our decoupled load-balancing framework:
\begin{itemize}
    \item \textbf{SC (Synchronous Compute, Baseline, Paradigm 1 in Section~\ref{subsec:dp-design_analysis}):} Represents the naive practice (e.g., DDP for DP and \texttt{All-Gather} for TP). It employs naive non-partitioning (DP-SC) and synchronous collective communication (TP-SC), resulting in redundant computation and blocking synchronization.
    \item \textbf{NV-layerwise (Baseline, Paradigm 2 in Section~\ref{subsec:dp-design_analysis}):} NVIDIA's \texttt{layerwise\_optimizer} implementation. It assigns optimizer states at the granularity of whole layers to respect tensor boundaries. However, as analyzed in Section~\ref{sec:dp-method} and Appendix~\ref{subsec:layerwise-geometric-conflict}, this approach introduces a Geometric Incompatibility between task assignment and parameter partition. This mismatch precludes efficient bucket-based \texttt{Reduce-Scatter} overlapping and necessitating expensive \texttt{All-Reduce} operations.
    \item \textbf{ASC (Asynchronous Compute, Paradigm 3 in Section~\ref{subsec:dp-design_analysis}):} Adopts our proposed decoupled architecture to enable asynchronous execution (DP-ASC + TP-ASC) but utilizes naive partitioning without load-aware scheduling.
    \item \textbf{LB-ASC (Load-Balance Asynchronous Compute, Our Framework's Core Strategy):} The complete implementation of our proposed strategies (DP-LB-ASC + TP-LB-ASC), which integrates the $\alpha$-Balanced Static Partitioning for DP and the Greedy LPT scheduling for TP to minimize stragglers.
\end{itemize}

\begin{figure}[htbp]
    \centering
    \includegraphics[width=0.95\linewidth]{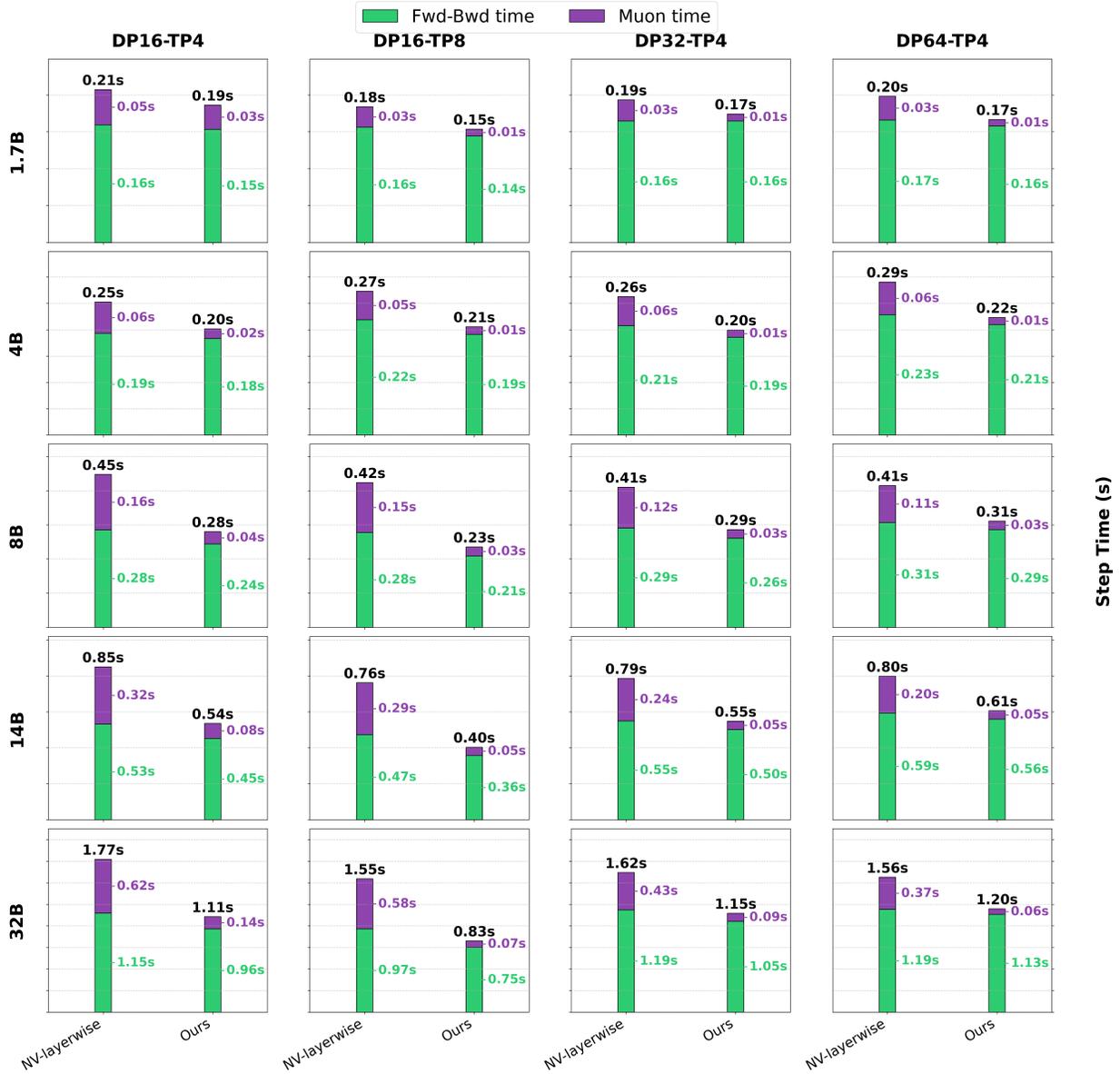}
    \caption{End-to-End Performance Comparison with NVIDIA \texttt{layerwise\_optimizer}. We compare the breakdown of step latency between \texttt{layerwise\_optimizer} (NV-layerwise) and our proposed framework across the Qwen3 model family (ranging from 1.7B to 32B parameters) under various parallelism configurations. (Top Row) Results for smaller models (1.7B, 4B) showing consistent latency reduction. (Bottom Row) Results for larger models (14B, 32B) where the computational complexity of the optimizer increases. Our method demonstrates robust scalability, with the performance gap widening as the model size and optimizer complexity grow, primarily due to the significant reduction in Optimizer Time (Purple bar).}
    \label{fig:full-compare-nv}
\end{figure}

\paragraph{Evaluation Metrics.}
We analyze system performance using two primary metrics:
\begin{enumerate}
    \item \textbf{Runtime Breakdown:} We report the wall-clock time for different training phases. 
    To provide a comprehensive performance context, we include the \textit{Fwd-Bwd time} (Forward and Backward pass) and standard \textit{AdamW optimizer time} as references. 
    The primary metric for comparison is the \textit{Target Optimizer time} (e.g., Muon step latency). All reported timings are averaged over 10 runs and 10 steps within each run to ensure stability.
    
    \item \textbf{Load-Balance Ratio:} To quantify the severity of the long-tail straggler problem, we define the Load-Balance Ratio ($R_{LB}$) for both memory footprint and computational FLOPs:
    \begin{equation}
        R_{LB} = \frac{\max_{r}(\mathcal{v}_r)}{\text{avg}_{r}(\mathcal{v}_r)}
    \end{equation}
    where $\mathcal{v}_r$ represents the metric value (Peak Memory or FLOPs) on rank $r$. An ideal balanced system yields $R_{LB} \approx 1.0$, while a higher ratio indicates severe imbalance.
\end{enumerate}

\subsection{Full Performance Comparison with \texttt{layerwise\_optimizer}}
\label{subsec:extended-exp-compare-nv}

To rigorously evaluate the superiority of our proposed framework over the current state-of-the-art solution for atomicity preservation, we conduct a comprehensive comparison against NVIDIA's \texttt{layerwise\_optimizer} on a cluster with up to 256 GPUs. Figure~\ref{fig:full-compare-nv} presents the step latency breakdown across the entire Qwen3 model family (1.7B, 4B, 8B, 14B, and 32B) under various Data Parallel (DP) and Tensor Parallel (TP) configurations.

\paragraph{Consistent Superiority.} The results demonstrate that our proposed framework consistently outperforms NV-layerwise in every tested configuration. The total iteration time reduction ranges from significant to dramatic, driven primarily by the optimization of the optimizer step (Purple bar). For instance, in the Qwen3-32B (DP16-TP8) setting, our method reduces the specific optimizer latency by approximately 8.3x compared to NV-layerwise.

\paragraph{Robustness to Model Size.} A key observation is that the performance gap widens as the model size increases. For smaller models like Qwen3-1.7B, the absolute time savings are noticeable but bounded by the relatively low computational cost of the optimizer. However, as we scale to Qwen3-32B, the matrix-based operations become computationally dominant. NV-layerwise, constrained by expensive collective \texttt{All-Reduce} and extra \texttt{All-gather}, exposes this latency fully. In contrast, our asynchronous pipeline effectively hides these expensive communication operators, resulting in a more pronounced speedup for larger models.

\paragraph{Robustness to Parallelism Strategies.} We further observe that our advantage remains robust regardless of the specific parallelism split (e.g., comparing DP16-TP8 vs. DP32-TP4). While NV-layerwise shows sensitivity to the communication patterns induced by different TP sizes, our static partitioning and micro-group scheduling adapt to the changing topology, maintaining high throughput. This confirms that our load-balancing algorithms are agnostic to the specific dimension of parallelism employed.

\begin{figure}[h]
    \centering
    \includegraphics[width=0.99\linewidth]{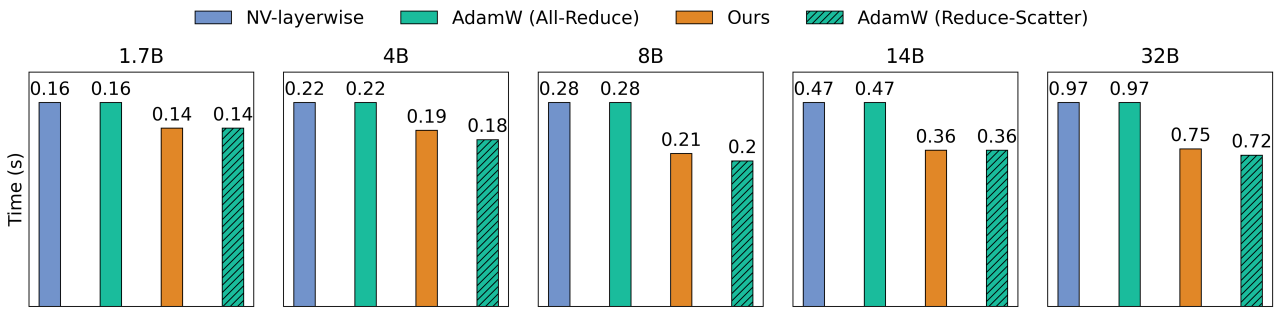}
    \caption{\textbf{Verification of Communication Efficiency in Forward-Backward Pass.} We compare the Fwd-Bwd step latency of our framework and the NV-layerwise against two controlled baselines: AdamW with \texttt{All-Reduce} (simulating the heavy communication volume necessitated by layer-wise partitioning) and \texttt{Reduce-Scatter} (representing the optimal ZeRO-1 communication volume). The results show that the NV-layerwise (blue) consistently aligns with the AdamW \texttt{All-Reduce} baseline (teal solid), confirming that its performance is bottlenecked by the 2x communication volume of \texttt{All-Reduce} operations despite overlapping. Conversely, our method (orange) closely tracks the AdamW \texttt{Reduce-Scatter} baseline (teal hatched), confirming that our static partitioning strategy successfully preserves the efficient communication capabilities of the underlying Megatron framework.}
    \label{fig:fwd-bwd-comm-overlap}
\end{figure}

\paragraph{Supplementary Experiment of Fwd-Bwd Communication Efficiency}
To quantitatively verify the effectiveness of communication overlap during the Forward-Backward (Fwd-Bwd) pass, we designed a comparative experiment. Since isolating precise communication and computation times within the complex pipelining of Fwd-Bwd is challenging, we established two reference baselines using the standard AdamW optimizer training in Megatron: (1) AdamW (\texttt{All-Reduce}, DDP), where gradients are collected via overlapping \texttt{All-Reduce}, representing the latency upper bound; and (2) AdamW (\texttt{Reduce-Scatter}, ZeRO-1), where \texttt{Reduce-Scatter} communication is optimally overlapped with computation (standard ZeRO-1 behavior), representing the theoretical lower bound.

Figure~\ref{fig:fwd-bwd-comm-overlap} demonstrates that our proposed framework successfully preserves the efficient bucket-based communication mechanism, achieving Fwd-Bwd latencies that closely track the AdamW (\texttt{Reduce-Scatter}) baseline. In contrast, the \texttt{layerwise\_optimizer} exhibits latencies matching the AdamW (\texttt{All-Reduce}) baseline, confirming that its violation of ZeRO-1 geometric constraints prevents effective hiding of communication overhead. We note that in certain configurations, our Fwd-Bwd time is marginally higher than the ideal AdamW (\texttt{Reduce-Scatter}) setting. This minor discrepancy arises because our load-balancing strategy necessitates variable-sized parameter chunks (Section~\ref{subsec:dp-system-workflow}), introducing slight communication imbalances compared to AdamW's perfectly equal chunks. However, this impact is negligible compared to the significant performance penalty of the inefficient \texttt{All-Reduce} primitive.

\subsection{Scaling Analysis}
\label{subsec:extended-exp-scaling-analysis}

\begin{figure}[h]
    \centering
    \begin{subfigure}[b]{0.9\textwidth}
        \centering
        \includegraphics[width=\linewidth]{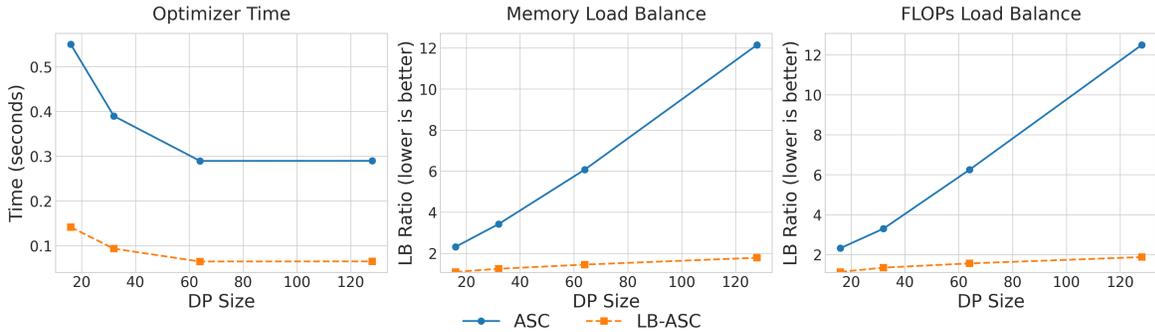}
        \caption{Data Parallelism Size Scaling Analysis (Fix TP=4)}
        \label{fig:dp-scaling}
    \end{subfigure}
    
    \par\bigskip
    
    \begin{subfigure}[b]{0.9\textwidth}
        \centering
        \includegraphics[width=\linewidth]{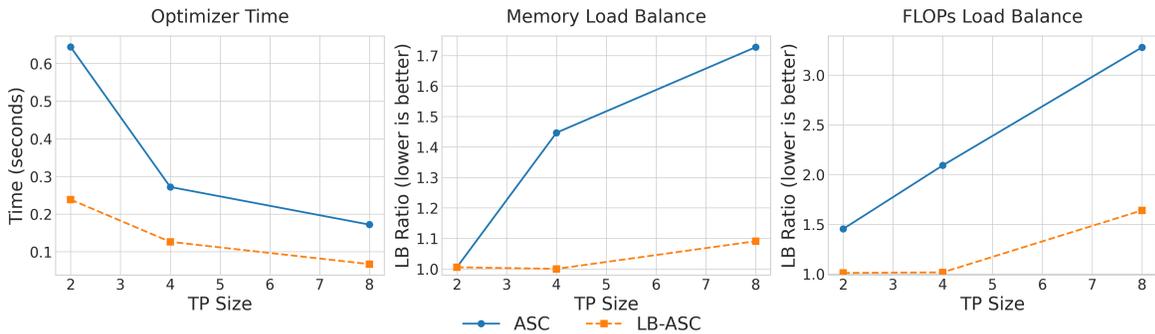}
        \caption{Tensor Parallelism Size Scaling Analysis (Fix PP=4 and DP=4)}
        \label{fig:tp-scaling}
    \end{subfigure}
    
    \caption{\textbf{Parallelism Scaling Analysis.}
    We evaluate the scalability of our load balancing algorithms as the degree of parallelism increases. (a) Data Parallelism Scaling: We fix the model (Qwen3-32B, Muon) and scale DP size from 16 to 128. While our proposed ASC suffers from increasing load imbalance (rising load-balance ratio) and slower convergence, our $\alpha$-Balanced strategy (DP LB-ASC) maintains a near-optimal load-balance ratio ($\approx 1.0$) and stable optimizer time. (b) Tensor Parallelism Scaling: Similarly, scaling TP size from 2 to 8 reveals that our Micro-Group Scheduling (TP LB-ASC) effectively neutralizes the straggler effect that plagues the non-load-balanced ASC.}
    \label{fig:parallelism-scaling}
\end{figure}

\begin{figure}[h]
    \centering
    \begin{subfigure}[b]{0.85\textwidth}
        \centering
        \includegraphics[width=\linewidth]{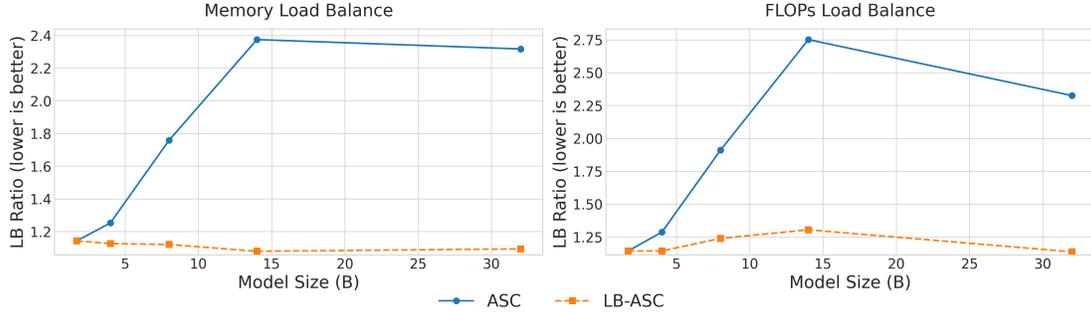}
        \caption{Model Size Scaling with DP-load-balance Analysis}
        \label{fig:model-size-scaling-dp-lb}
    \end{subfigure}
    
    \par\bigskip
    
    \begin{subfigure}[b]{0.85\textwidth}
        \centering
        \includegraphics[width=\linewidth]{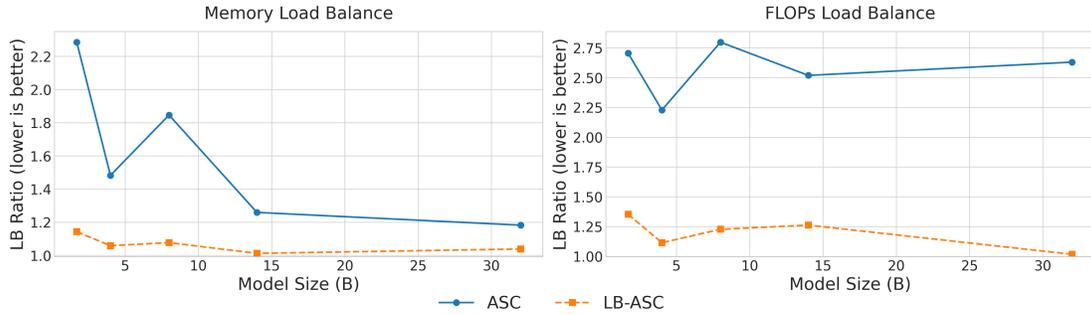}
        \caption{Model Size Scaling with TP-load-balance Analysis}
        \label{fig:model-size-scaling-tp-lb}
    \end{subfigure}
    
    \caption{\textbf{Model Size Scaling Analysis.}
    We evaluate the robustness of our load balancing algorithms across different model sizes (1.7B to 32B). (a) Data Parallelism: Larger models introduce greater parameter heterogeneity, causing the load-balance ratio of the baseline (ASC) to degrade significantly. Our LB-ASC strategy maintains a consistent balance. (b) Tensor Parallelism: The load imbalance fluctuates based on the specific hidden dimension alignment, yet our Greedy Scheduling consistently optimizes the packing.}
    \label{fig:model-size-scaling}
\end{figure}

We investigate the scalability of our approach on a large-scale cluster of up to 512 GPUs from two dimensions: system size (Parallelism Scaling) and workload complexity (Model Size Scaling).

\paragraph{Parallelism Scaling (Figure~\ref{fig:parallelism-scaling}).} We fix the model size (Qwen3-32B) and scale the number of GPUs. 
\begin{itemize}
    \item \textbf{Data Parallelism (Figure~\ref{fig:dp-scaling}):} With TP fixed at 4, as we scale the DP size from 16 to 128, the statistical variance in parameter sizes naturally leads to severe load imbalance in naive strategies. The ASC baseline exhibits a linear degradation in the load-balance ratio for both memory and FLOPs, leading to increased optimizer time due to stragglers. Conversely, our DP LB-ASC ($\alpha$-Balanced) strategy maintains a load-balance ratio close to the ideal 1.0, effectively neutralizing the straggler effect even at large scale (128 DP ranks).
    \item \textbf{Tensor Parallelism (Figure~\ref{fig:tp-scaling}):} Similarly, with PP fixed at 4 and DP fixed at 4, increasing TP size typically exacerbates the fragmentation of weight matrices. The baseline shows a sharp increase in computational imbalance. Our Micro-Group Scheduling (TP LB-ASC) successfully mitigates this, keeping the FLOPs load-balance ratio significantly lower than the baseline, thereby preserving low optimizer latency as the system scales.
\end{itemize}

\paragraph{Model Size Scaling (Figure~\ref{fig:model-size-scaling}).} We analyze how parameter heterogeneity affects load balancing as models grow from 1.7B to 32B. For these experiments, we fix the parallelism configuration to DP=16 and TP=4.
\begin{itemize}
    \item \textbf{DP Load Balance (Figure~\ref{fig:model-size-scaling-dp-lb}):} Larger models typically contain a wider variance of tensor shapes (e.g., larger embedding layers vs. smaller projection heads). This heterogeneity causes the naive partitioning's load-balance ratio to increase significantly with model size. Our LB-ASC strategy adapts to this heterogeneity, maintaining a flat LB profile.
    \item \textbf{TP Load Balance (Figure~\ref{fig:model-size-scaling-tp-lb}):} Interestingly, for TP, the baseline imbalance does not strictly increase with model size but fluctuates based on the specific architecture (e.g., hidden dimension size). However, our greedy scheduling algorithm consistently finds a near-optimal packing, ensuring that our performance advantage is sustained across all model sizes.
\end{itemize}

\subsection{Generalization to Other Optimizers (Shampoo \& SOAP)}
\label{subsec:extended-exp-generalization-to-other-optim}

\begin{figure}[h]
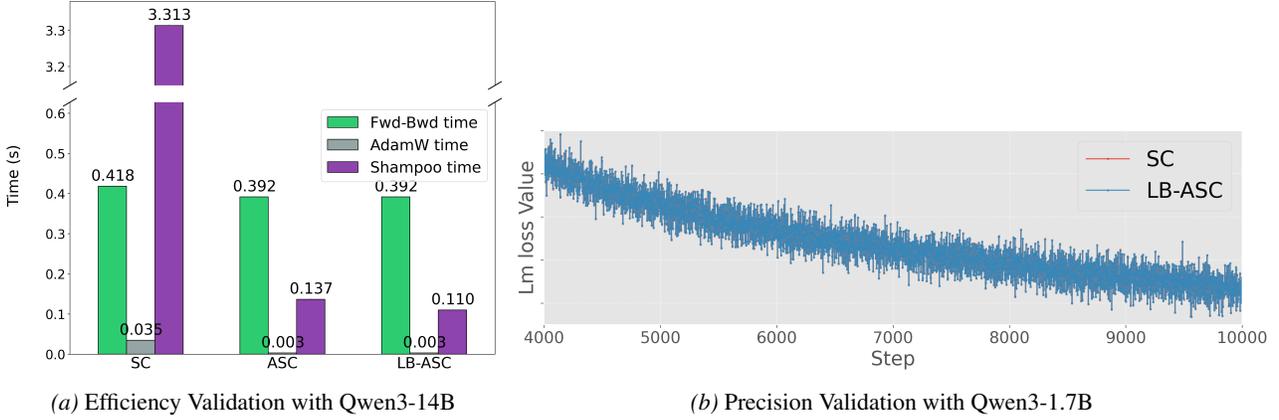

    \centering
    \begin{subfigure}[b]{0.39\textwidth}
        \centering
        \includegraphics[width=\linewidth]{figures/exp-efficiency-shampoo.pdf}
        \caption{Efficiency Validation with Qwen3-14B}
        \label{fig:validation-shampoo-efficiency}
    \end{subfigure}
    \begin{subfigure}[b]{0.59\textwidth}
        \centering
        \includegraphics[width=\linewidth]{figures/exp-prec-shampoo-qwen3-1.7B-dp8-tp4.pdf}
        \caption{Precision Validation with Qwen3-1.7B}
        \label{fig:validation-shampoo-precision}
    \end{subfigure}
    \caption{\textbf{Generality Validation with Shampoo Optimizer.} (a) Efficiency: Comparison of step time for Qwen3-14B using the Shampoo optimizer. The results demonstrate that our proposed LB-ASC strategy is equally effective for Shampoo, achieving a drastic speedup compared to both the synchronous (SC) and asynchronous (ASC) methods. (b) Precision: Training loss curves for Qwen3-1.7B confirm that our system optimizations do not alter the convergence trajectory compared to the synchronous baseline.}
    \label{fig:validation-shampoo}
\end{figure}

\begin{figure}[h]
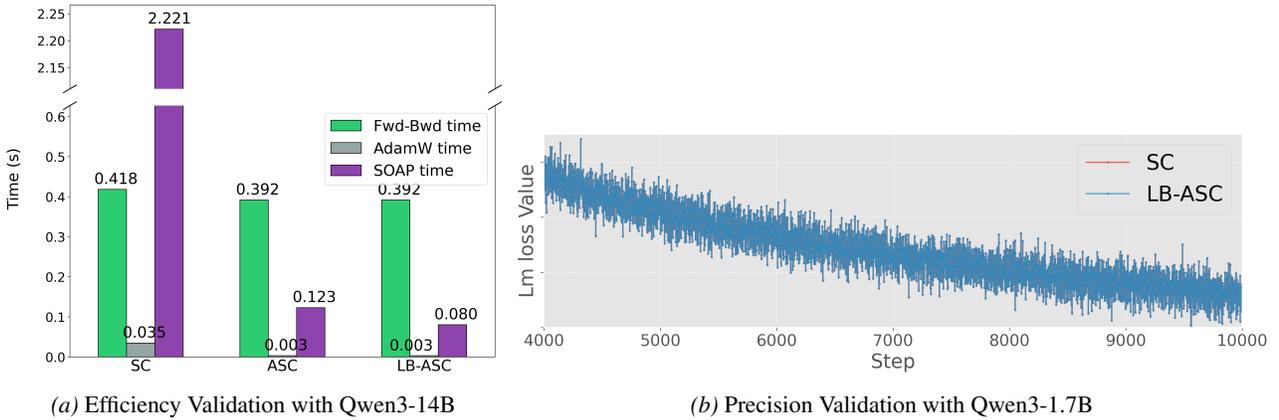

    \centering
    \begin{subfigure}[b]{0.39\textwidth}
        \centering
        \includegraphics[width=\linewidth]{figures/exp-efficiency-soap.pdf}
        \caption{Efficiency Validation with Qwen3-14B}
        \label{fig:validation-soap-efficiency}
    \end{subfigure}
    \begin{subfigure}[b]{0.59\textwidth}
        \centering
        \includegraphics[width=\linewidth]{figures/exp-prec-soap-qwen3-1.7B-dp8-tp4.pdf}
        \caption{Precision Validation with Qwen3-1.7B}
        \label{fig:validation-soap-precision}
    \end{subfigure}
    \caption{\textbf{Generality Validation with SOAP Optimizer.} (a) Efficiency: Step time comparison for Qwen3-14B trained with the SOAP optimizer. Similar to the Shampoo results, our framework generalizes well to SOAP, significantly reducing step latency. (b) Precision: The loss curve for Qwen3-1.7B confirms that our implementation introduces no algorithmic deviation, ensuring mathematical equivalence with standard synchronous training.}
    \label{fig:validation-soap}
\end{figure}

To verify the generality of our framework beyond Muon, we evaluate its performance on Shampoo and SOAP, two other representative matrix-based optimizers. These algorithms involve distinct cubic-complexity operations (SVD and Eigendecomposition, respectively), serving as robust test cases for our system's adaptability.

\paragraph{Efficiency (Figures~\ref{fig:validation-shampoo-efficiency} \& \ref{fig:validation-soap-efficiency}).} These experiments are conducted on 256 GPUs. Due to the significant memory overhead of Shampoo and SOAP preconditioners, we utilize Qwen3-14B with a configuration of PP=2, DP=32, and TP=4 to accommodate the state requirements. The matrix operations in Shampoo and SOAP incur high computational costs, as evidenced by the high latency in the SC baseline (e.g., Shampoo step takes 3.313s). The experimental results confirm that our proposed framework remains highly effective in this context. By applying our unified partitioning and scheduling strategy, we reduce the optimizer step time to 0.110s for Shampoo and similarly for SOAP. This represents a speedup of over 30x, demonstrating that our framework can successfully handle diverse matrix-based workloads without requiring algorithm-specific tuning.

\paragraph{Precision (Figures \ref{fig:validation-shampoo-precision} \& \ref{fig:validation-soap-precision}).} We validate correctness by training a Qwen3-1.7B model for 400B tokens using a DP=8, TP=4 configuration. The loss curves for both Shampoo and SOAP under our framework overlap perfectly with the standard synchronous baseline. It is important to emphasize that LB-ASC strategy is a purely system-level optimization. Unlike methods that approximate curvature or skip updates to save time, our approach strictly adheres to the original optimizer's mathematical definition. Consequently, there is zero mathematical precision loss, and the convergence behavior is identical to that of the baseline.

\begin{figure}[h]
    \centering
    \begin{subfigure}[b]{0.99\textwidth}
        \centering
        \includegraphics[width=\linewidth]{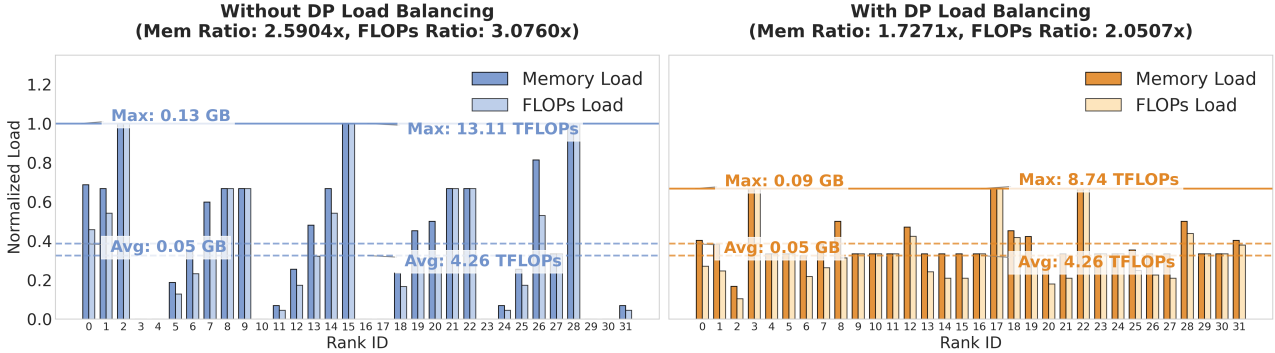}
        \caption{Data Parallelism Load Balance Ratio}
    \end{subfigure}
    
    \par\bigskip
    
    \begin{subfigure}[b]{0.6\textwidth}
        \centering
        \includegraphics[width=\linewidth]{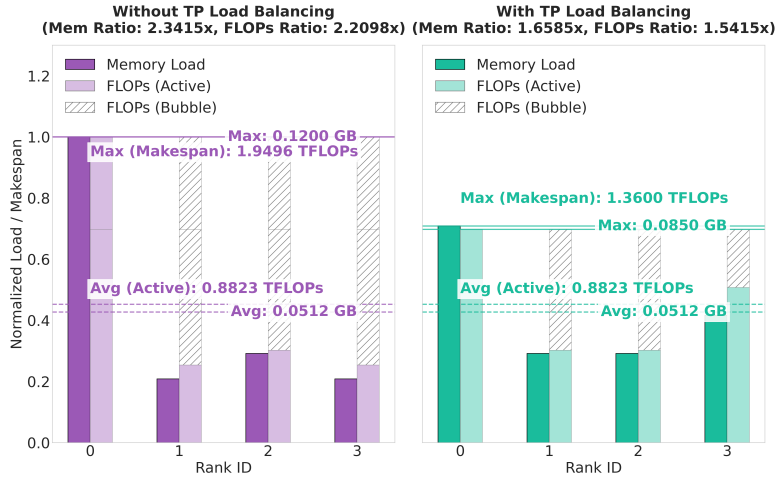}
        \caption{Tensor Parallelism Load Balance Ratio}
    \end{subfigure}
    
    \caption{\textbf{Load Balance Analysis for Shampoo/SOAP.} Visualization of the load distribution (Memory and FLOPs) across ranks for the Shampoo optimizer. Similar to the Muon results, the naive partitioning (Left) leads to significant computational bubbles and memory peaks, while our proposed load-balancing strategy (Right) effectively flattens the distribution, achieving a near-perfect load-balance ratio.}
    \label{fig:pp2-dp32-tp4-lb-ratio}
\end{figure}

\paragraph{Load Balance (Figure~\ref{fig:pp2-dp32-tp4-lb-ratio}).} The load distribution analysis for Shampoo / SOAP further corroborates the effectiveness of our proposed LB-ASC strategy. Despite the different computational characteristics of Shampoo compared to Muon, our scheduler successfully flattens the workload variance (reducing the FLOPs load-balance ratio from $> 2.0$ to $\approx 1.05$), ensuring efficient hardware utilization across all ranks.

\subsection{DP Load Balance $\alpha$ Analysis}
\label{subsec:extended-exp-dp-alpha}

\begin{figure}[h]
    \centering
    \includegraphics[width=0.8\linewidth]{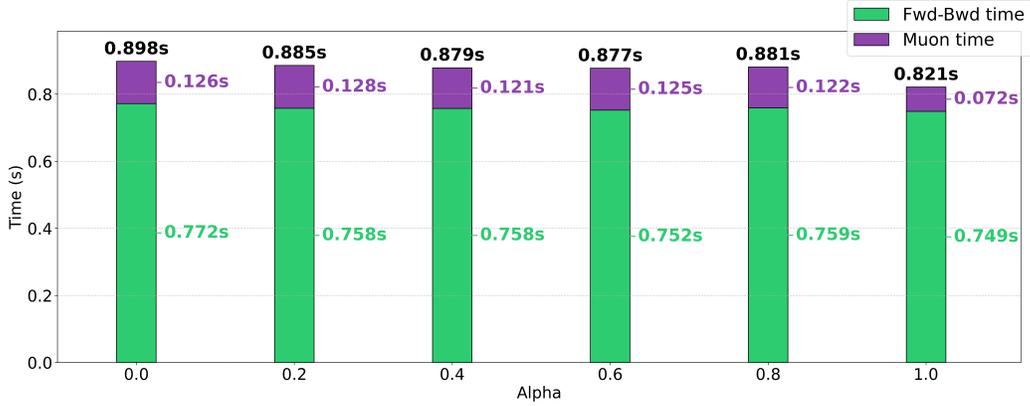}
    \caption{\textbf{Sensitivity to DP Load-Balance Factor $\alpha$.}
    We analyze the impact of the control parameter $\alpha$ (ranging from 0.0 to 1.0) on the total step latency for Qwen3-32B. $\alpha=0$ prioritizes communication size uniformity, while $\alpha=1$ prioritizes optimizer computational load balance. The results indicate that setting $\alpha=1.0$ yields the lowest total time, suggesting that computational stragglers in the optimizer step are the dominant bottleneck, while communication imbalance in the Fwd-Bwd pass is effectively hidden by overlap.}
    \label{fig:dp-alpha}
\end{figure}

We conduct an ablation study on 128 GPUs (configured as PP=8, DP=16) to determine the optimal value of the control parameter $\alpha$ in our $\alpha$-Balanced Greedy LPT algorithm (Section 3.2). Recall that $\alpha \to 0$ prioritizes uniform communication sizes (matching standard ZeRO), while $\alpha \to 1$ prioritizes balanced optimizer computational load. As shown in Figure~\ref{fig:dp-alpha}, as $\alpha$ increases from 0.0 to 1.0, the Muon time (Purple bar) decreases monotonically. This confirms that computational stragglers are the primary bottleneck during the optimizer step. Crucially, while increasing $\alpha$ introduces slight non-uniformity in the communication bucket sizes (potentially affecting Fwd-Bwd time), the Fwd-Bwd time (Green bar) remains relatively stable. This stability suggests that the efficient communication overlapping in Megatron effectively masks the minor communication imbalances introduced by our sizing strategy. Consequently, setting $\alpha = 1.0$ yields the best end-to-end performance, justifying our design choice to prioritize computational load balancing in the final implementation.

\subsection{TP Communication Fusion Analysis}
\label{subsec:extended-exp-tp-fuse-size}

We analyze the impact of the Micro-Group Fusion strategy for Tensor Parallelism (Section 4.1) using 128 GPUs with DP=16 and TP=8. We vary the maximum capacity constraint ($C_{max}$) of the micro-groups, which dictates the buffer size for the fused All-to-All communication.

\begin{figure}[h]
    \centering
    \includegraphics[width=0.7\linewidth]{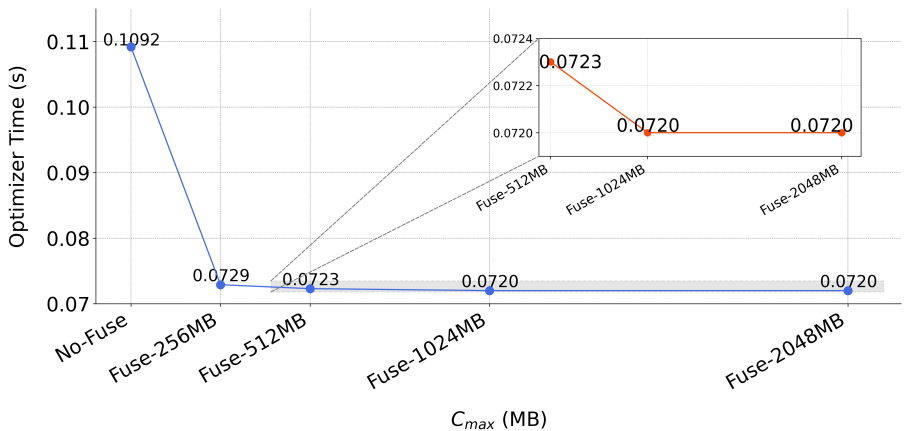}
    \caption{\textbf{TP Micro-Group Fusion.}
    We evaluate the impact of fusing tensor updates into Micro Groups with varying capacity constraints ($C_{max}$). "No-Fuse" represents treating each tensor individually. The results demonstrate that fusing communication significantly reduces optimizer time by saturating All-to-All bandwidth. Performance plateaus once $C_{max}$ exceeds 1024MB, indicating the optimal granularity for overlap.}
    \label{fig:tp-fuse-size}
\end{figure}

Figure~\ref{fig:tp-fuse-size} illustrates the Optimizer Time as a function of $C_{max}$. The ``No-Fuse'' baseline, which performs communication for each tensor individually, suffers from high latency ($\approx 0.11s$) due to the overhead of launching many small kernels and poor bandwidth utilization. Enabling fusion immediately drops the latency to $\approx 0.073s$. We observe that performance improves slightly as $C_{max}$ increases, stabilizing around 512MB. Beyond this point (1024MB, 2048MB), the latency plateaus, indicating that the All-to-All bandwidth is fully saturated. This result confirms that our fusion strategy is essential for efficiency and that the system is robust to the specific choice of $C_{max}$ provided it is sufficiently large to saturate the interconnect.

\section{Supplementary Explanation}

\subsection{Load-Balance Algorithm Latency}

It is worth noting that the load-balancing algorithms described in Section~\ref{subsec:dp-load-balance-optimization} and Section~\ref{subsec:tp-balance} are executed strictly as a \textbf{one-time offline planning step} during model initialization. Consequently, they introduce \underline{almost zero overhead} to the subsequent training iterations. Furthermore, the computational cost of this initialization phase itself is negligible: both algorithms are designed as efficient greedy heuristics with a time complexity dominated by sorting ($O(N \log N)$), where $N$ is the number of parameter tensors or buckets (typically in the scale of thousands).

In our large-scale experiments with Qwen3-32B on 256 GPUs, the offline planning phase completes in milliseconds. This latency is orders of magnitude smaller than the heavy system initialization procedures, such as CUDA context creation, memory allocation, and NCCL communicator setup (which typically span tens of seconds to minutes). Consequently, the scheduling overhead is effectively imperceptible and cannot be distinguished from system noise during the model initialization stage.

Given that this overhead is effectively imperceptible and indistinguishable from system noise, we omit specific latency experiments for the load-balancing phase in our reported results.

\subsection{The Geometric Incompatibility of Layer-wise Partitioning}
\label{subsec:layerwise-geometric-conflict}

\begin{figure}[htbp]
    \centering
    \includegraphics[width=0.98\linewidth]{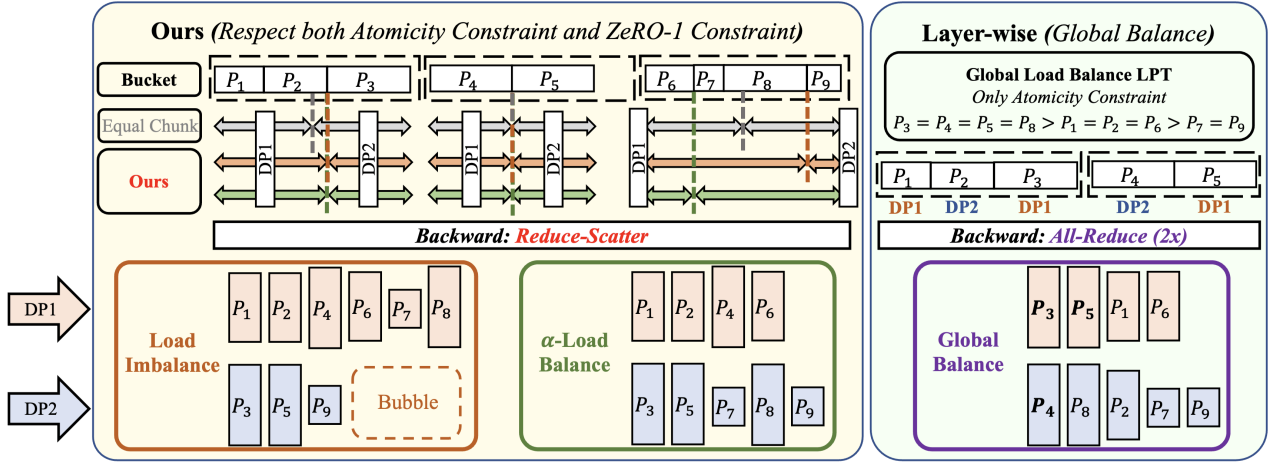}
    \caption{\textbf{Visualization of Geometric Incompatibility.} 
    \textbf{(Left) Ours:} Our approach respects both Atomicity and the strict Geometric Constraints of ZeRO-1 primitives. By optimizing slice sizes (shifting boundaries) inside bucket without scrambling the physical parameter sequence, we preserve the contiguous bucket abstraction. This alignment enables the use of the optimal \texttt{Reduce-Scatter} primitive.
    \textbf{(Right) Layer-wise:} Strategies relying on Global Load Balancing (e.g., Global LPT) assign tasks purely by weight (e.g., assigning heavy parameter $P_3$ to DP1) regardless of their physical location. This \underline{Data-Task Mismatch} (visualized by the interleaved DP1/DP2 assignments in the bucket view) breaks bucket coalescing. Consequently, under Megatron ZeRO-1’s fixed geometric slicing and bucket coalescing assumption, the system must abandon bucket-based \texttt{Reduce-Scatter} in favor of bucket-based \texttt{All-Reduce} or per-parameter-based \texttt{Reduce-Scatter}, incurring \underline{2x communication volume} and higher latency. (Note: The visualization of the third bucket ($P_6, P_7, P_8, P_9$) in the right sub-figure is omitted here because its load-based assignment ($P_6 \in \text{DP1}, \{P_7, P_8, P_9\} \in \text{DP2}$) coincidentally aligns with the geometric rank order (monotonic), thus not violating the ZeRO-1 Geometric Constraint.)}
    \label{fig:geometric-conflict}
\end{figure}

\textbf{Mechanism (The "Hybrid" ZeRO-1).} 
Existing layer-wise strategies (e.g., NVIDIA's \texttt{layerwise\_optimizer}) operate as a hybrid system to reconcile atomicity with distribution. During the Forward and Backward passes, they function in a DDP mode and utilize standard \texttt{All-Reduce} (incurring $2\times$ communication volume compared to ZeRO-1's Reduce-Scatter) to synchronize gradients. However, for the Optimizer Step, they shard the optimizer states and employ a \textbf{Global Load Balancing} strategy (Global LPT) to assign the ownership of optimizer states. This assignment is based strictly on computational weights (e.g., placing heavy layers on specific ranks) to equalize execution time, disregarding the physical memory layout of the parameters.

\textbf{The Conflict (ZeRO Geometric Incompatibility in Megatron).} 
This design introduces a fundamental conflict between the logical task assignment and the physical communication primitives used in modern training stacks like ZeRO-1.

\begin{itemize}
    \item \textbf{ZeRO-1 Primitive (Position-Based):} The efficient \texttt{Reduce-Scatter} primitive (Appendix~\ref{sec:extended-preliminary}) relies on a strict \textbf{Geometric Partitioning} logic (Figure~\ref{fig:geometric-conflict} (Right)). For a contiguous bucket of size $B$ across $R$ ranks, the primitive rigidly dictates that the $r$-th geometric slice (interval $[\frac{r-1}{R}B, \frac{r}{R}B)$) is sent to Rank $r$. The destination is determined solely by the data's \textit{physical position} in the buffer.
    
    \item \textbf{Layer-wise Logic (Weight-Based):} In contrast, the layer-wise allocator assigns parameter update tasks based on \textit{load}. As illustrated in Figure~\ref{fig:geometric-conflict} (Right), a heavy parameter such as $P_3$ might be assigned to \textbf{DP1} to balance the global workload, even though its physical position in the buffer falls within a region that geometrically belongs to or is interleaved with \textbf{DP2}. This creates a clear \textbf{Data-Task Mismatch}.
\end{itemize}

\textbf{The Consequence: The "Lose-Lose" Dilemma.}
Due to this data-task mismatch, layer-wise approaches are forced to choose between two suboptimal communication paths:

\begin{itemize}
    \item \textbf{Option A: The Bandwidth Penalty (Forced All-Reduce).} 
    This is the default choice for current \texttt{layerwise\_optimizer} implementations. Since the system cannot coalesce the geometrically-fragmented task assignments into a single geometry-based \texttt{Reduce-Scatter}, it defaults to \texttt{All-Reduce}. This broadcasts full gradients to all ranks, ensuring that every rank receives the necessary data regardless of its assignment. However, this incurs \textbf{2x the communication volume} compared to ZeRO-1's \texttt{Reduce-Scatter}, significantly increasing the bandwidth requirement.
    
    \item \textbf{Option B: The Latency Penalty (Forced Reduce-Scatter).} 
    Alternatively, one might attempt to enforce a \texttt{Reduce-Scatter} (sending only the specific parameter data to its assigned rank). However, this fundamentally breaks the \textbf{Bucket Coalescing} abstraction. 
    Because the destination ranks are no longer geometrically aligned—visualized in Figure~\ref{fig:geometric-conflict} as the interleaved assignment sequence $P_1(\text{DP1}) \rightarrow P_2(\text{DP2}) \rightarrow P_3(\text{DP1})$—the system cannot launch a single monolithic kernel for the entire bucket. Instead, it must dismantle the bucket and trigger \textbf{discrete, per-parameter communication kernels} to route specific tensors to their arbitrary destinations. This introduces massive kernel launch overhead and prevents the system from saturating the interconnect bandwidth, shifting the bottleneck from bandwidth to latency.
\end{itemize}

\textbf{Compounded Penalty on Parameter Update.}
Crucially, this geometric violation extends beyond gradient synchronization to parameter reconstruction. Under ZeRO-1 geometric constraints, the updated parameters are gathered via a bucket-based \texttt{All-Gather} that is efficiently overlapped with the Forward pass computation. However, because layer-wise partitioning misaligns the ownership of the updated parameters with their geometric shards, a standard coalesced \texttt{All-Gather} is impossible. Consequently, implementations are forced to either: (1) perform an explicit \texttt{All-Gather} or \texttt{Broadcast} \textit{within} the Optimizer Step to redistribute weights (adding significant exposed latency, as observed in our baseline experiments); or (2) resort to inefficient per-parameter \texttt{All-Gather} operations during the Forward pass, which destroy the overlap efficiency.

In summary, our framework outperforms layer-wise approaches because our $\alpha$-Balanced Partitioning respects not only the \textbf{Atomicity Constraint} but also the \textbf{Geometric Constraints of ZeRO-1 primitives}. By optimizing slice sizes without breaking the sequential rank ordering, we retain the efficient, coalesced \texttt{Reduce-Scatter} capability in backward, and the corresponding overlapped, coalesced \texttt{All-Gather} in forward, that layer-wise strategies are structurally forced to abandon.

\subsection{Detailed Walkthrough of Load-Balancing Algorithms}
\label{subsec:alg_walkthrough}

\begin{algorithm}[h]
   \caption{Micro-Group Construction with Greedy Rollback}
   \label{alg:tp_scheduling}
\begin{algorithmic}[1]
   \STATE {\bfseries Input:} Parameters $\mathcal{P}$, Shapes $\mathcal{S}$, Capacity $C_{max}$, Ranks $R$
   \STATE {\bfseries Output:} List of Micro Groups $\mathbb{M}_{opt}$ where each $M \in \mathbb{M}_{opt}$ contains assignments $\{(p, rank)\}$
   
   \STATE \COMMENT{Phase 1: Deterministic Global Sort}
   \STATE $\mathcal{I}_{meta} \leftarrow [(\text{Cost}(p), p, s) \text{ for } p, s \text{ in } \text{zip}(\mathcal{P}, \mathcal{S})]$
   \STATE Sort $\mathcal{I}_{meta}$ descending by Cost
   
   \STATE $\mathbb{M}_{opt} \leftarrow []$
   \STATE $M_{curr} \leftarrow []$ \COMMENT{Current candidate items}
   \STATE $idx \leftarrow 0$

   \WHILE{$idx < \text{Length}(\mathcal{I}_{meta})$}
       \STATE $item \leftarrow \mathcal{I}_{meta}[idx]$
       \STATE $M_{curr}.\text{append}(item)$
       
       \STATE \COMMENT{Simulation Step: Try to solve partition for current candidate list}
       \STATE $Assignments, L_{max} \leftarrow \textsc{MinHeapSolver}(M_{curr}, R)$
       
       \IF{$L_{max} \le C_{max}$}
           \STATE \COMMENT{Valid: Item fits. Continue accumulating.}
           \STATE $idx \leftarrow idx + 1$
       \ELSE
           \STATE \COMMENT{Constraint Violated: Trigger Rollback}
           \STATE $M_{curr}.\text{pop}()$ \COMMENT{Remove the item that caused overflow}
           
           \IF{$M_{curr}$ is empty}
                \STATE \textbf{Error:} Single item exceeds $C_{max}$.
           \ENDIF
           
           \STATE \COMMENT{Finalize the previous valid group}
           \STATE $FinalAlloc, \_ \leftarrow \textsc{MinHeapSolver}(M_{curr}, R)$
           \STATE $\mathbb{M}_{opt}.\text{append}(FinalAlloc)$
           
           \STATE \COMMENT{Start new group with the current item}
           \STATE $M_{curr} \leftarrow []$
           \STATE \COMMENT{Do not increment idx; retry item in next iteration}
       \ENDIF
   \ENDWHILE
   
   \STATE \COMMENT{Handle remaining items}
   \IF{$M_{curr}$ is not empty}
       \STATE $FinalAlloc, \_ \leftarrow \textsc{MinHeapSolver}(M_{curr}, R)$
       \STATE $\mathbb{M}_{opt}.\text{append}(FinalAlloc)$
   \ENDIF
   
   \STATE \textbf{Return} $\mathbb{M}_{opt}$
\end{algorithmic}
\end{algorithm}

\begin{algorithm}[h]
   \caption{Subroutine: MinHeapSolver (LPT)}
   \label{alg:minheap_solver}
   \begin{algorithmic}[1]
   \STATE {\bfseries Input:} Items $I$, Ranks $R$
   \STATE {\bfseries Output:} Allocation $\mathcal{A}$, MaxLoad $L_{max}$
   
   \STATE Sort $I$ descending by Cost \COMMENT{Local LPT Sort}
   \STATE PriorityQueue $PQ \leftarrow [(0, r) \text{ for } r \in 0..R-1]$ \COMMENT{Min-Heap of (load, rank)}
   \STATE $\mathcal{A} \leftarrow [[] \text{ for } r \in 0..R-1]$
   
   \FOR{each item $(c, p, s)$ in $I$}
       \STATE $min\_load, r_{best} \leftarrow \text{heapq.heappop}(PQ)$
       \STATE $\mathcal{A}[r_{best}].\text{append}((p, s))$
       \STATE $new\_load \leftarrow min\_load + c$
       \STATE $\text{heapq.heappush}(PQ, (new\_load, r_{best}))$
   \ENDFOR
   
   \STATE $L_{max} \leftarrow \max_{(l, r) \in PQ}(l)$
   \STATE \textbf{Return} $\mathcal{A}, L_{max}$
   \end{algorithmic}
\end{algorithm}

In this section, we provide a rigorous walkthrough of the two core scheduling algorithms proposed in the paper. We explain the mathematical intuition behind the variable definitions and the algorithmic decisions.

\paragraph{Algorithm~\ref{alg:greedy_lpt}: \texorpdfstring{$\alpha$}{alpha}-Balanced Static Partitioning (DP)}

The core challenge in DP is determining the cut points $s_{i,r}$ for a bucket $B_i$ such that we balance two objectives: \textbf{DP Computational Load} vs. \textbf{Fwd-Bwd Bucket Communication Uniformity}.

\textbf{Intuition of the Control Parameter $\alpha$:}
The algorithm constructs a \textit{Target Allocation Vector} $v^*$ for the current bucket by blending two basis vectors:
\begin{itemize}
    \item \textbf{$v_{fill}$ (The History Basis):} Represents a compensatory strategy. It assigns more work to ranks that are currently ``underloaded'' (have a load deficit) based on previous allocations.
    \item \textbf{$v_{even}$ (The Geometry Basis):} Represents a perfectly uniform partition based on data size ($1/R$). This corresponds to the standard ZeRO-1 strategy.
\end{itemize}

The parameter $\alpha \in [0, 1]$ controls this blend. If $\alpha=1$, the algorithm aggressively fills deficits (prioritizing compute balance). If $\alpha=0$, it ignores history (prioritizing communication uniformity).

\textbf{Step-by-Step Execution:}
\begin{enumerate}
    \item \textbf{Global Sorting (LPT Rule):} We process buckets in descending order of their total cost $\mathcal{W}$. This ``Longest Processing Time'' (LPT) rule is crucial because large rocks (heavy buckets) are harder to balance and should be placed when the ``deficit voids'' are most flexible.
    \item \textbf{Deficit Calculation:} At step $t$, we calculate the current load vector $L$. The mean load $\mu$ represents the ideal average. The deficit vector $d = \max(0, \mu - L_r)$ identifies which ranks are lagging behind the ideal average.
    \item \textbf{Target Construction:} We calculate a ``fill proportion'' vector $v_{fill} = d / \sum d$, which dictates how much of the current bucket should be given to each rank to help them catch up. The final target proportion is $v^* = (1-\alpha)v_{even} + \alpha v_{fill}$.
    \item \textbf{Atomic Discretization:} The target vector $v^*$ gives us a continuous ``ideal'' load. We iterate through the parameter list and place the cut point $u$ strictly at parameter boundaries to minimize the distance to the target cumulative sum, thereby enforcing the \textbf{Atomicity Constraint}.
\end{enumerate}

\paragraph{Algorithm~\ref{alg:tp-balance}: Micro-Group Scheduling with Greedy Rollback (TP)}

For TP, we must partition a stream of tensor updates into discrete \underline{Micro Groups} (for fused communication) and assign them to \underline{Host Ranks} (for computation). This is modeled as a Bin Packing Problem with an embedded Multiprocessor Scheduling Problem. Based on our implementation, we define a \underline{Two-Level Greedy Strategy} detailed in Algorithm~\ref{alg:tp_scheduling} and Algorithm~\ref{alg:minheap_solver}.

\textbf{Algorithmic Logic Explanation:}

\begin{enumerate}
    \item \textbf{Global vs. Local Sorting:}
    We perform a \underline{Global Sort} (Line 5, Alg~\ref{alg:tp_scheduling}) of all parameters first. This ensures that we attempt to pack the largest, most difficult tensors into groups first, preventing ``straggler tensors'' from accumulating at the end. Inside the \textsc{MinHeapSolver} (Line 3, Alg~\ref{alg:minheap_solver}), we perform a \underline{Local Sort} as part of the standard Longest Processing Time (LPT) heuristic, which provides a reliable approximation for makespan minimization.

    \item \textbf{The Simulation \& Rollback Mechanism:}
    The core innovation is the feedback loop between the Bin Packing (Outer loop) and the Scheduling (Inner solver). Instead of estimating the cost of a group by simple summation ($\sum Cost / R$), we run the \underline{actual scheduling simulation} (\textsc{MinHeapSolver}) at every step (Line 13, Alg~\ref{alg:tp_scheduling}). This calculates the \textit{exact} makespan $L_{max}$ given the current items. If adding a new item causes $L_{max}$ to exceed $C_{max}$ (due to fragmentation or imbalance), we strictly enforce the limit by triggering a \underline{Rollback} (Lines 18-26). The current group is finalized, and the overflow item becomes the seed for the next group.

    \item \textbf{Efficiency:}
    Although simulating the solver at every step seems computationally intensive, the \textsc{MinHeapSolver} operates in $O(K \log R)$ for $K$ items. Since the number of items per Micro Group ($K$) is small (typically $< 50$), this overhead is negligible during the offline planning phase.
\end{enumerate}

\subsection{Discussion on the Choice of Training Framework}
\label{subsec:choice-of-training-framework}

In the realm of large-scale distributed training, \textbf{Megatron} \cite{shoeybi2019megatron} and \textbf{PyTorch FSDP} (Fully Sharded Data Parallel) \cite{zhao2023pytorch} represent the two dominant frameworks. In this work, we chose to implement and evaluate our strategies primarily within Megatron for two key reasons: (1) it remains the de facto standard for training massive-scale foundational models (e.g., GPT-3, Qwen) due to its highly optimized tensor parallelism and pipelining capabilities; and (2) its reliance on \underline{strict geometric partitioning} provided the ideal testbed to demonstrate how our Static Partitioning \underline{respects ZeRO-1 Geometric Constraints} to preserve coalesced communication overlapping, a feature structurally abandoned by other atomicity-preserving methods.

However, the principles of our proposed framework, particularly the \textbf{Load-Balanced Asynchronous Compute (LB-ASC)}, are not limited to Megatron and can be generalized to FSDP. From a high-level system abstraction, the optimizer step in FSDP shares fundamental structural similarities with the Tensor Parallelism (TP) challenge we address in Section~\ref{sec:tp-method}. 

Specifically, both FSDP and TP fragment parameters across devices. To perform a matrix-based optimizer update (which requires strict atomicity), both paradigms necessitates an \texttt{All-Gather} operation to reconstruct the full parameter, followed by the computation, and a shard update. Therefore, our \underline{Micro-Group Scheduling} and \underline{Asynchronous Compute pipeline} are directly transferable to FSDP. By treating FSDP shards as logical units similar to TP splits, our load-balancing algorithm can similarly effectively neutralize stragglers and maximize throughput in an FSDP-based training stack.

It is crucial to note that adapting our TP LB-ASC strategy to FSDP requires distinct handling depending on the specific ZeRO stage employed. 

\begin{itemize}
    \item \textbf{FSDP-ZeRO1 Scenario:} In this configuration, gradients are typically not sharded (they are replicated across ranks within the sharding group). Since our TP LB-ASC primarily serves as a gradient reconstruction mechanism, the \texttt{All-to-All} communication depicted in Figure~\ref{fig:async-tp} can be skipped.
    
    \item \textbf{FSDP-ZeRO2/3 Scenario:} Here, gradients are sharded across devices. Consequently, the full execution flow of our TP LB-ASC (including the fused \texttt{All-to-All} reconstruction) is necessary within the optimizer step to assemble the required data.
\end{itemize}

\noindent \textbf{Communication Overhead and Engineering Complexity in Hybrid Parallelism:} 
Our choice is driven by the distinct communication characteristics of Tensor Parallelism (TP) versus Data Parallelism (DP). TP typically leverages high-bandwidth intra-node interconnects (e.g., NVLink); thus, while parameter sharding is unavoidable, performing reconstruction communication within the TP group during the optimizer step is acceptable. In contrast, DP communication usually spans lower-bandwidth inter-node networks, making any synchronization within the DP group prohibitively expensive. (Notably, this bottleneck constitutes the primary drawback of the \texttt{layerwise\_optimizer} baseline, which necessitates an expensive \texttt{All-Reduce} and an additional \texttt{All-Gather} operation over these slow links during the step.) A significant challenge arises when superimposing TP onto FSDP-ZeRO2/3: this hybrid setting necessitates a coupled 2D (DP+TP) communication mesh to reconstruct tensors fragmented across both dimensions. This not only forces expensive synchronization over the slower inter-node DP links but also introduces substantial engineering complexity to manage the hybrid topology. Megatron's architecture avoids this by inherently decoupling DP and TP, enabling us to isolate synchronization strictly within the high-speed TP domain while keeping the DP dimension communication-free (via ZeRO-1).

\begin{wrapfigure}{r}{0.25\textwidth}
    \vspace{-15pt}
    \centering
    \includegraphics[width=1.0\linewidth]{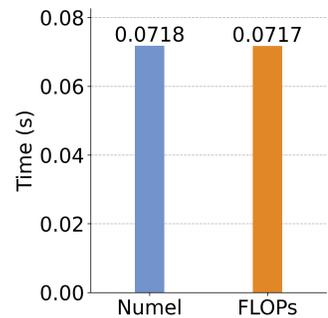}
    \caption{\textbf{Cost Metric Ablation.} Comparison of Muon step latency using Numel vs. exact FLOPs cost function. The difference is negligible ($\approx 10^{-4}$s).}
    \label{fig:cost-function}
    \vspace{-15pt}
\end{wrapfigure}

\subsection{Discussion about Load-Balance  Generalization with Non-linear Cost}
\label{sec:non-linear-generalization}

While our framework mathematically supports arbitrary cost functions $\mathcal{W}(p)$ to handle the cubic complexity of matrix-based optimizers, in our practical implementation and experiments, we adopted the simplified linear metric $\mathcal{W}(p) = \text{numel}(p)$ primarily to enforce the Unified design philosophy of our framework.We justify this choice based on three key rationales:
\begin{itemize}
    \item Optimizer-Agnostic Universality: A core contribution of this work is decoupling the system architecture from the specific optimization algorithm. By using numel as the universal cost metric, our partitioner remains strictly agnostic to the optimizer's internal complexity (whether $\mathcal{O}(N)$ or $\mathcal{O}(N^3)$), ensuring that the same infrastructure seamlessly supports Muon, Shampoo, SOAP, and future algorithms without modification.
    \item Shape-Cost Correlation: For standard Transformer architectures (e.g., Qwen), parameter tensors typically fall into grouped shapes where computational cost correlates strongly with parameter count, making numel an effective proxy.
    \item Implementation Robustness: This simplification avoids the engineering fragility of maintaining distinct cost models for every new optimizer, reducing system complexity while retaining high load-balancing efficiency as demonstrated in our experiments.
\end{itemize}

To quantitatively validate the effectiveness of this simplification, we conducted a controlled experiment using the Qwen3-32B model with a parallelism configuration of DP=16 and TP=8. 

We compared the execution makespan of the optimizer step when scheduling with the exact computational complexity ($\mathcal{W}(p)=\text{FLOPs}$) versus the simplified parameter count ($\mathcal{W}(p)=\text{numel}(p)$). 

As illustrated in Figure~\ref{fig:cost-function}, the performance difference is negligible: using FLOPs yields a latency of \textbf{0.0717s}, while using Numel yields \textbf{0.0718s}. The discrepancy is approximately $10^{-4}$s. In the context of our broader results in Section~\ref{sec:extended-exp}, where the performance gaps between different methods typically range from $10^{-2}$s to $10^{-1}$s, this minute deviation confirms that the shape-based cost metric ($\text{numel}$) serves as a highly accurate proxy for load balancing. It captures the dominant workload characteristics of Transformer blocks without introducing the engineering overhead of calculating exact FLOPs for every tensor.

\section{Extended Related Work}
\label{sec:extended-related-work}

To situate our contributions, we classify existing distributed optimization strategies into three categories: (1) \textbf{System-Level Compromises} (e.g., Layer-wise Partitioning), which preserve mathematical exactness but sacrifice pipeline efficiency; (2) \textbf{Algorithmic Approximations} (e.g., Blocking/Shard-local), which sacrifice mathematical fidelity to fit system constraints; and (3) our proposed Canzona, which reconciles both without compromise.

\subsection{The Fundamental Conflict: Matrix-based Optimizers vs. Distributed Atomicity}

Modern large-scale training relies on \underline{Geometric Sharding} (e.g., ZeRO-1, Tensor Parallelism (\S~\ref{sec:extended-preliminary})) to distribute memory. Frameworks like Megatron flatten parameters into a contiguous \texttt{param\_and\_grad\_buffer} and slice them uniformly (e.g., into $1/N$ shards) regardless of tensor boundaries.

\begin{itemize}
    \item \textbf{The Conflict:} Matrix-based optimizers (e.g., 
    Muon~\citep{jordan6muon}, 
    Shampoo~\citep{gupta2018shampoo}, 
    SOAP~\citep{vyas2024soap}, 
    Conda~\citep{wang2025conda}, 
    ROOT~\citep{he2025root}, 
    PSGD~\citep{li2017preconditioned}, 
    Sophia~\citep{liu2023sophia}, 
    K-FAC~\citep{martens2015optimizing, osawa2019large}, 
    and \citep{xie2026controlled}) are inherently non-separable. They require holistic access to the full weight matrix $W \in \mathbb{R}^{d_{in} \times d_{out}}$ to perform operations like SVD or Newton-Schulz iterations.
    \item \textbf{The Gap:} Standard geometric sharding violates this \underline{Atomicity Constraint}. A single weight matrix is often fragmented across multiple ranks, making local execution of these mathematical operations impossible without costly reconstruction.
\end{itemize}

\subsection{System-Level Attempts: Exactness at the Cost of Efficiency}

To preserve the mathematical exactness of the optimizer, recent industrial implementations have adopted coarser-grained partitioning.

\paragraph{Layer-wise Partitioning (e.g., NVIDIA's \texttt{layerwise.optimizer}, Distributed Shampoo~\citep{shi2023distributed}).} 
This approach assigns optimizer states at the granularity of whole layers to ensure atomicity. While mathematically correct, it introduces severe system-level bottlenecks driven by the Geometric Incompatibility (detailed in \S\ref{subsec:layerwise-geometric-conflict}):

\begin{enumerate}
    \item \textbf{Structural Fallback to \texttt{All-Reduce}:} Because the logical assignment conflicts with the geometric slicing of ZeRO primitives, the system cannot coalesce parameters into a single \texttt{Reduce-Scatter} kernel. It is structurally forced to fallback to \texttt{All-Reduce} operations ($2\times$ communication volume) and an additional \texttt{Broadcast}/\texttt{All-Gather} step during the optimizer update to synchronize the new parameters across ranks (as shown in Figure~\ref{fig:comparison_nv} and \ref{fig:full-compare-nv}).
\end{enumerate}

\subsection{Algorithmic Attempts: Efficiency at the Cost of Fidelity}

To avoid the heavy communication of exact reconstruction, another line of research modifies the optimizer algorithm itself, introducing approximations.

\paragraph{Block-Diagonal Approximations (Small-Block Methods).} 
Methods like Distributed Shampoo~\citep{anil2020scalable} and K-FAC~\citep{osawa2019large} approximate the full preconditioner with a block-diagonal matrix (e.g., approximating a $4096 \times 4096$ correlation with smaller blocks).
\begin{itemize}
    \item \textbf{Limitation:} This ignores off-diagonal correlations. While effective for smaller models or specific architectures (e.g., CNNs), this ``Small-Block'' assumption may degrade convergence speed or solution quality for Large Language Models where dense correlations matter.
\end{itemize}

\paragraph{Shard-Local Orthogonalization (e.g., MuonBP).} 
Recent proposals like MuonBP~\citep{khaled2025muonbp} attempt to bypass global communication by performing orthogonalization strictly on the local shard of the weight matrix.
\begin{itemize}
    \item \textbf{Limitation:} This constitutes a \underline{Mathematical Approximation}. The ``Local Newton-Schulz'' update differs from the ``Global Newton-Schulz'' update. This creates a \underline{Directional Drift} between the true gradient geometry and the applied update, which can lead to training instability or require frequent, expensive synchronization steps to correct.
\end{itemize}

\paragraph{Low-Rank and Subspace Approximations.} 
To circumvent the prohibitive cost of full-rank matrix operations, a recent wave of optimizers proposes constraining the update trajectory to lower-dimensional manifolds to reduce communication and memory overheads. 
\textbf{Dion}~\citep{ahn2025dion} introduces a distributed orthonormalization technique that replaces global Newton-Schulz iterations with decoupled momentum buffers and low-rank approximations, allowing states to evolve independently on shards without expensive synchronization. 
Similarly, methods utilizing dynamic subspace selection, such as ~\citep{gong2025towards}, project second-order information onto predefined orthogonal bases of low-rank structures. 
\begin{itemize}
    \item \textbf{Limitation:} While these approaches drastically improve efficiency, they impose strong structural assumptions—effectively assuming weight updates lie strictly in a low-rank subspace. This constitutes a fundamental \textit{fidelity trade-off}: unlike system-level attempts, which support the exact holistic update required by the original algorithm, these methods alter the optimization path, potentially degrading convergence quality in regimes where full-rank curvature information is critical.
\end{itemize}

\paragraph{Overall Limitation:} Beyond the potential degradation in convergence fidelity and the need for further verification at scale, these algorithmic approximations fundamentally lack a Unified design philosophy. Instead of offering a general system solution, they often resort to ad-hoc, algorithm-specific modifications—such as tailoring strategies strictly for Muon (e.g., MuonBP)—which restricts their extensibility to other emerging or future matrix-based optimizers.

\subsection{Our Solution: Canzona (System-Level Exact, Efficient, and Unified Optimization)}

Our framework, \textbf{Canzona}, distinguishes itself by simultaneously achieving \underline{Zero-Fidelity-Loss}, high system throughput, and broad universality. Unlike prior works that compromise on one or more of these axes, our framework delivers a comprehensive solution:

\begin{itemize}
    \item \textbf{System-Level Exactness (vs. Algorithmic Approximations):} We strictly adhere to the original mathematical definition of the optimizer. By handling the atomicity constraint purely through memory layout and scheduling, rather than altering the update rule (e.g., via blocking or shard-local approximations), we guarantee Zero-Fidelity-Loss, ensuring convergence behaviors identical to single-device baselines.

    \item \textbf{Straggler-Free Efficiency (vs. Layer-wise Partitioning):} We address the system bottlenecks across both parallelism dimensions. For Data Parallelism, our $\alpha$-Balanced Static Partitioning preserves the computation balance during the optimizer update step. By successfully reconciling the Atomicity Constraint with ZeRO-1 Geometric Constraints, it allows us to fully inherit the efficient, \underline{coalesced communication overlapping} of the standard ZeRO-1 pipeline and zero communication during optimizer step. Simultaneously, for Tensor Parallelism, our Asynchronous Micro-Group Scheduling eliminates the redundancy of synchronous execution. By batching fragmented updates and balancing computational loads, we effectively hide the expensive reconstruction overhead that limits existing layer-wise approaches.

    \item \textbf{Unified and Agnostic Framework (vs. Specific Hacks):} Crucially, our framework is designed as a Unified framework decoupled from specific optimizer logic. We treat tensor updates as generic computational tasks defined solely by cost metrics (e.g., numel). This makes our system Optimizer-Agnostic, allowing it to support not only Muon but also Shampoo, SOAP, and future matrix-based algorithms without requiring algorithm-specific "hacks" or re-engineering.
\end{itemize}

\end{document}